\documentclass[11pt,a4paper,logo]{googledeepmind}
\setleftlogo[95pt]{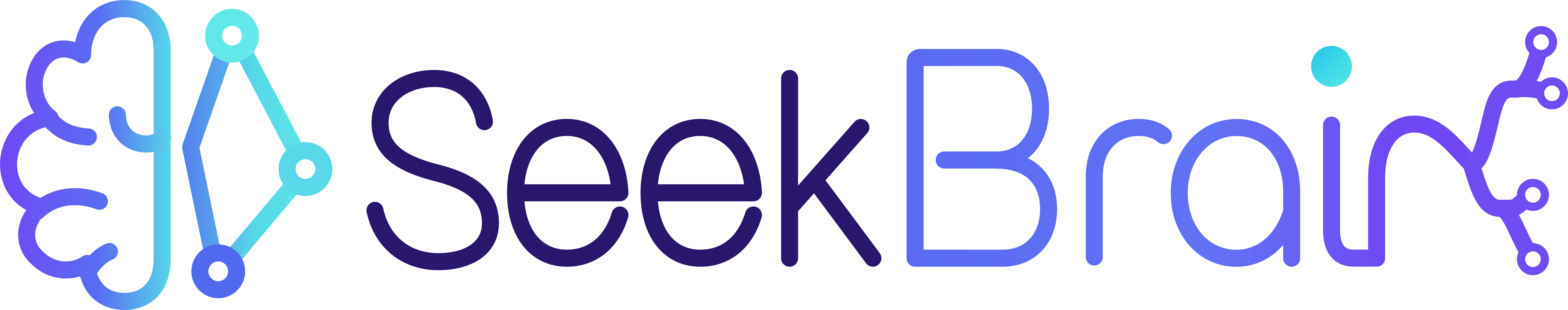} 
\setrightlogo[160pt]{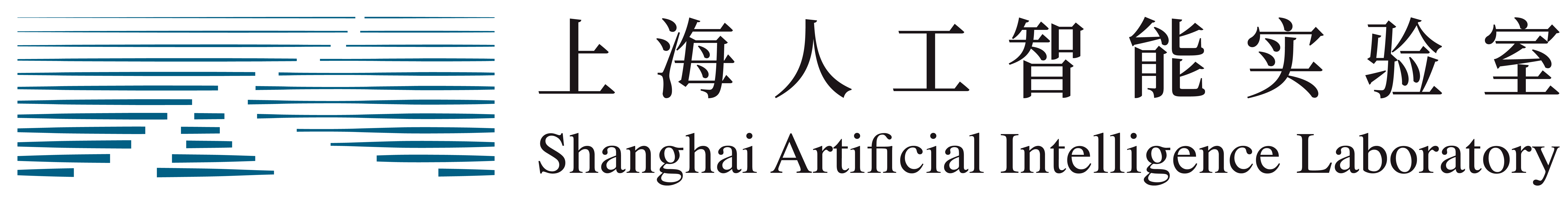}
\usepackage{fontawesome5}
\usepackage{marvosym}
\usepackage{academicons}
\usepackage{etoolbox}
\usepackage[T1]{fontenc}
\usepackage{tgcursor}
\usepackage{pifont}
\usepackage[
    natbib=true,
    backend=biber,
    style=numeric, 
    sorting=none, 
    maxbibnames=5, 
    minbibnames=5  
]{biblatex}
\usepackage{seqsplit}

\makeatletter
\renewcommand\AB@authnote[1]{}   
\renewcommand\AB@affilnote[1]{}  
\makeatother

\AtEveryBibitem{\clearfield{month}}
\AtEveryBibitem{\clearfield{day}}
\usepackage{csquotes}

\title{\includegraphics[height=0.72cm]{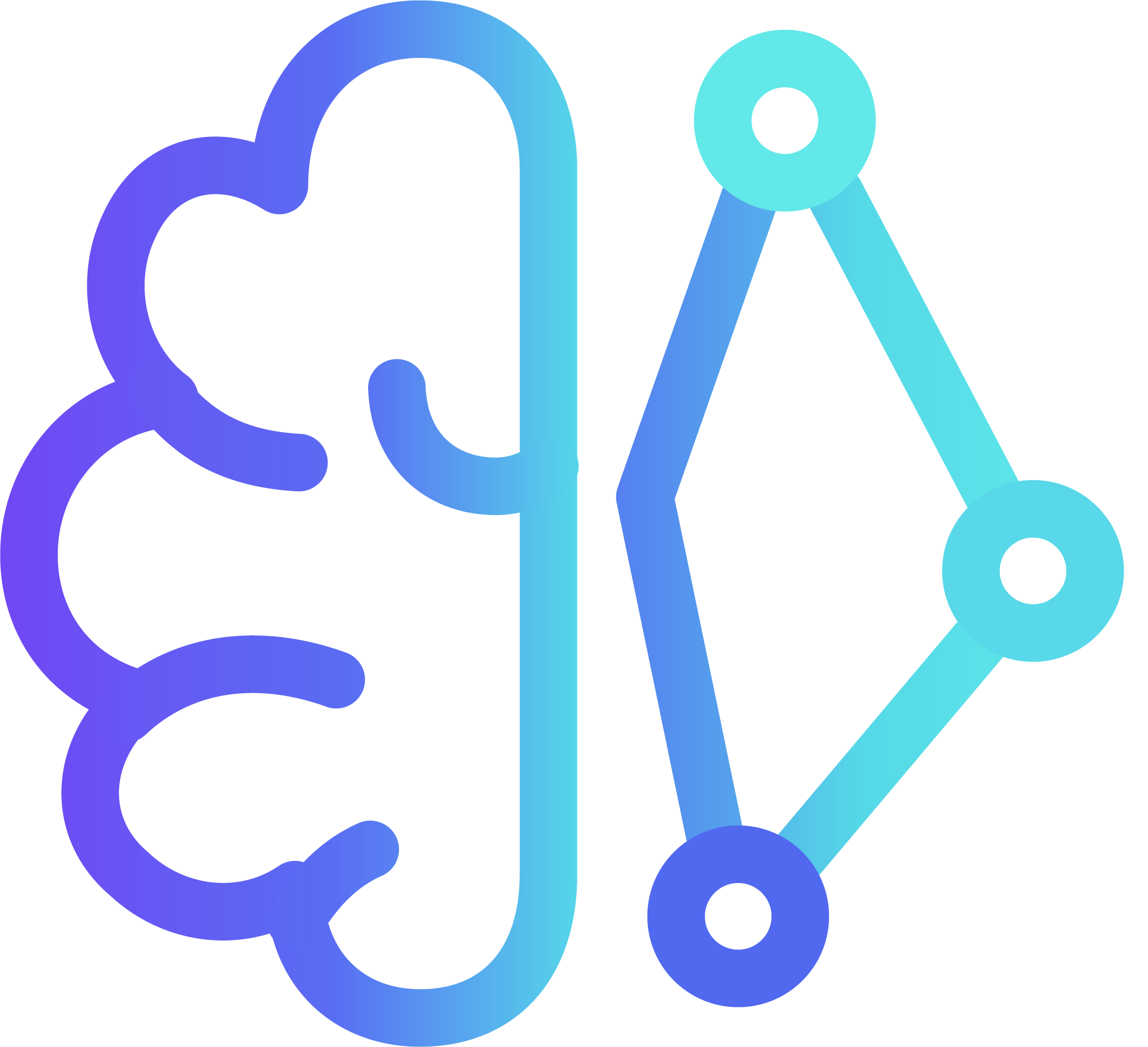}\hspace{0.1em} {\fontsize{20.5}{22.5}\selectfont SeekBrain: An Autonomous Multi-Agent System for Accelerating Neuroscience Discovery}}

\renewcommand\Affilfont{\normalfont\fontsize{9}{11}\selectfont}

\makeatletter
\renewcommand\AB@affilsepx{\par\vspace{-4pt}\Affilfont}
\makeatother
\renewcommand\Affilfont{\normalfont\raggedright\fontsize{9}{11}\selectfont}   

\correspondingauthor{
\faEnvelope[regular]~{songchunfeng@pjlab.org.cn} 
}

\author{
Jiamin Wu\textsuperscript{1, 3~*}, 
Peishan Xiang\textsuperscript{1~*}, 
Jingyang Chen\textsuperscript{1~$\dag$}, 
Yuqing Zhu\textsuperscript{1~$\dag$}, 
Yuxi Li\textsuperscript{2~$\dag$},
Ling Luo\textsuperscript{1~$\dag$},
Qihao Zheng\textsuperscript{1~$\dag$},
Jialiang Zu\textsuperscript{2}, 
Yongchao Wu\textsuperscript{4}, 
Mindong Liu\textsuperscript{1}, 
Haitao Wu\textsuperscript{1}, 
Chaofan Hu\textsuperscript{1}, 
Yijie Sun\textsuperscript{1}, 
Yuqi Hang\textsuperscript{6},
Yu Zhu\textsuperscript{1},  
Shuo Li\textsuperscript{1}, 
Yue Fan\textsuperscript{1}, 
Shiyang Feng\textsuperscript{1}, 
Wanghan Xu\textsuperscript{1}, 
Tianlei Zhang\textsuperscript{2},
Jie Zhang\textsuperscript{5}, 
Wenlong Zhang\textsuperscript{1}, 
Bo Zhang\textsuperscript{1}, 
Kai Wang\textsuperscript{2}, 
Lei Bai\textsuperscript{1~$\spadesuit$}, 
Mianxin Liu\textsuperscript{1~\Letter$\spadesuit$}, 
Wanli Ouyang\textsuperscript{1, 3~\Letter}, 
Jiulin Du\textsuperscript{1, 2~\Letter}, 
Chunfeng Song\textsuperscript{1~\Letter$\spadesuit$}
}

\affil{\vspace{2pt}\textsuperscript{1}Shanghai Artificial Intelligence Laboratory,
\textsuperscript{2}State Key Laboratory of Brain Cognition and Brain-Inspired Intelligence Technology, Center for Excellence in Brain Science and Intelligence Technology, Chinese Academy of Sciences,
\textsuperscript{3}The Chinese University of Hong Kong,
\textsuperscript{4}Southeast University,
\textsuperscript{5}Fudan University,
\textsuperscript{6}New York University
} 

\affil{\vspace{6pt}$*$~ Co-first Author \quad
$\dag$~Core Contributor \quad 
$\spadesuit$~Project Lead \quad
\Letter~Corresponding Author 
} 

\usepackage{pdflscape}
\usepackage{textcomp}
\usepackage{rotating}
\usepackage{setspace}
\usepackage{microtype} 
\usepackage{graphicx}
\usepackage{tabularx}
\usepackage{subcaption}
\usepackage{caption}

\usepackage{booktabs}
\usepackage[table]{xcolor}
\usepackage{xcolor}

\usepackage{array}
\usepackage{threeparttable}
\usepackage{multirow}
\usepackage{bm}

\usepackage{amsmath}
\usepackage{siunitx}

\usepackage{enumitem}
\usepackage{float}
\usepackage{seqsplit}
\usepackage{framed}

\usepackage{tikz}

\usepackage{algorithm}
\usepackage{algpseudocode}
\usepackage{amssymb}
\usepackage{microtype}
\usepackage{xcolor}
\algrenewcommand\algorithmiccomment[1]{\hfill $\triangleright$ #1}
\algnewcommand\algorithmicforeach{\textbf{for each}}
\algdef{S}[FOR]{ForEach}[1]{\algorithmicforeach\ #1\ \algorithmicdo}
\usepackage{caption} 
\usepackage[export]{adjustbox}

\usepackage{ragged2e}
\usepackage[most]{tcolorbox}

\usepackage{makecell}
\usepackage{tabularray}

\usepackage{longtable}
\usepackage{listings}

\definecolor{CaseGreen}{HTML}{2E7D32} 
\definecolor{CaseOrange}{HTML}{F57C00} 
\definecolor{CaseGray}{HTML}{F6F7F8}   
\definecolor{CaseInk}{HTML}{212121}    
\definecolor{CaseWhite}{HTML}{F5F5DC}
\definecolor{DeepBlue}{HTML}{003366} 
\definecolor{LightBlue}{HTML}{99CCFF} 
\definecolor{DeepPurple}{HTML}{673AB7} 
\definecolor{MiddlePurple}{HTML}{9C7FD0}
\definecolor{LightPurple}{HTML}{D1C4E9} 
\definecolor{HotPink}{HTML}{FF69B4} 
\definecolor{SoftPink}{HTML}{F8BBD0} 
\definecolor{Crimson}{HTML}{DC143C} 
\definecolor{Teal}{HTML}{008080} 
\definecolor{Cyan}{HTML}{00BCD4} 
\definecolor{SoftGray}{HTML}{EEEEEE}       
\definecolor{LighterGray}{HTML}{FAFAFA} 

\newtcolorbox{stagebox}[1]{
  breakable, enhanced,
  colback=CaseGray, colframe=DeepBlue, coltitle=CaseWhite,
  title=\bfseries #1, fonttitle=\bfseries,
  left=1.2mm,right=1.2mm,top=1.2mm,bottom=1.2mm, boxrule=0.6pt
}

\newtcblisting{verbatimbox}{
  listing only, breakable, enhanced,
  colback=white, colframe=LightBlue, boxrule=0.5pt,
  left=1.2mm,right=1.2mm,top=1.2mm,bottom=1.2mm
}

\usepackage[symbol]{footmisc}
\newcolumntype{Y}{>{\RaggedRight\arraybackslash}X}

\usepackage{hyperref}
\hypersetup{
    colorlinks=true,
    linkcolor=blue, 
    citecolor=blue,  
    filecolor=black,
    urlcolor=blue    
}

\newcommand{\homepage}{\raisebox{-1.5pt}{\includegraphics[height=1em]{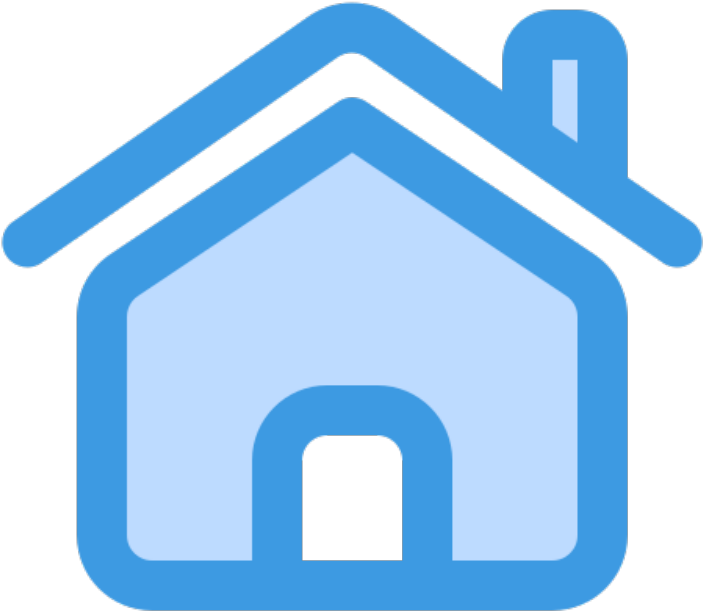}}}
\newcommand{\github}{\raisebox{-1.5pt}{\includegraphics[height=1em]{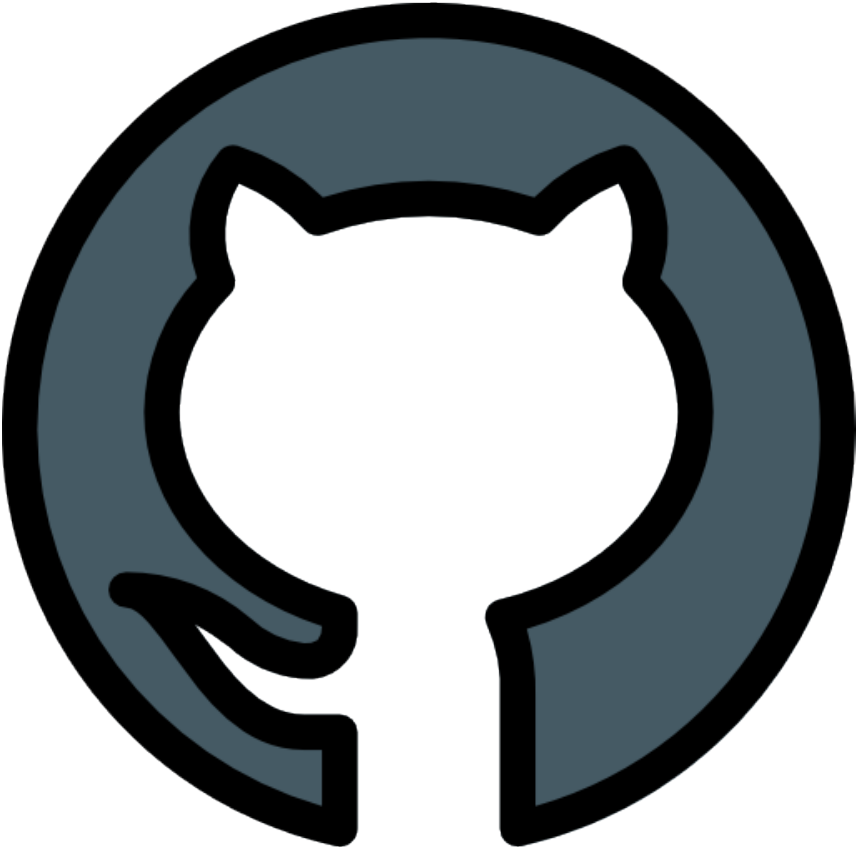}}}

\begin{abstract}

Modern neuroscience relies on integrating multi-scale, multimodal datasets to uncover the neural principles underlying intelligence. 
However, analytical challenges posed by highly heterogeneous data and fragmented workflows increasingly constrain discoveries. Here we introduce SeekBrain, an autonomous multi-agent framework designed to accelerate neuroscience discovery through domain-grounded hierarchical planning and cross-modal data analysis. 
SeekBrain dynamically constructs a repertoire of analysis recipes extracted from code-paper pairs. By coupling this codified expertise with agentic planning and execution engines, the framework scalably generates hypotheses and analytical pipelines on demand. Systematic evaluation on the expert-annotated BrainArena benchmark demonstrates that SeekBrain substantially outperforms state-of-the-art agent baselines across various analysis tasks. Crucially, when deployed in real-world research, SeekBrain integrated behavioral, neural, and anatomical data to reveal structured, distributed neural representations of larval zebrafish behavior and a shared axis of regional decoding strength across the brain in a mouse decision-making task. These results establish SeekBrain as a scalable and practical tool for accelerating data-driven discoveries in neuroscience.

\vspace{\baselineskip}

\homepage~Project Page:~\href{https://ai4neurolab.github.io/SeekBrain-page/}{https://ai4neurolab.github.io/SeekBrain-page/}

\github~Code:~\href{https://github.com/AI4NeuroLab/SeekBrain}{https://github.com/AI4NeuroLab/SeekBrain}

\end{abstract}

\begin{document}

\sloppy 
\maketitle

\section{Introduction}

Decoding biological intelligence is a central challenge in modern science.
To achieve this, neuroscience research must link multiple biological scales to uncover the mechanisms underlying neural computation.
Recent experimental technologies now allow us to observe the brain across molecular profiles, structural connectivity, neural dynamics and behavioral~\cite{biccn2021multimodal, finn2023functional, yao2023highresolution, microns2025functional, dorkenwald2024neuronal, ibl2025brainwide, arkhipov2025integrating, mathis2026joint}. 
Yet, despite the increasing availability of these high-dimensional datasets~\cite{biccn2021multimodal, yao2023highresolution, microns2025functional, dorkenwald2024neuronal, ibl2025brainwide}, progress is constrained by the complexity of analyzing them.  
Extracting neuroscientific insight from such diverse data requires specialized expertise, cross-disciplinary reasoning, and extensive trial and error. 
Consequently, routine data processing and coding consume much of the research cycle,
leaving many cross-scale hypotheses untested and valuable data underused~\cite{finn2023functional, arkhipov2025integrating, mathis2026joint}.
Accelerating discovery in neuroscience therefore demands automated frameworks that scale domain expertise to drive an iterative cycle of hypothesis exploration and rigorous data analysis.

To address similar bottlenecks in other scientific domains, Large Language Model (LLM)-driven agents have recently emerged to automate key research stages~\cite{robin, aiscientist, ai-coscientist, cellvoyager, software, feng2026internagent, team2025internagent, internagent-a1, Marwitz2026}. They have demonstrated the capability to generate testable biomedical hypothesis~\cite{ai-coscientist, Luo2025}, identify drug repurposing candidates~\cite{robin}, analyze single-cell RNA (scRNA) data~\cite{cellvoyager}, and optimize empirical scientific software~\cite{software}. 
AI-assisted discovery has primarily advanced in fields supported by robust computational tools and unified data representations. In bioinformatics, for instance, frameworks such as CellVoyager automates repetitive scRNA-seq analysis tasks by leveraging established software packages (\textit{e.g.}, scanpy~\cite{scanpy}) that encode decades of methodological expertise. Similarly, general-purpose scientific agents like Biomni~\cite{huang2025biomni} rely on clean, structured input data ready for direct computational analysis.

However, transferring these agent systems to neuroscience remains severely bottlenecked. 
The underlying challenge is a mismatch between current agent architectures and the fragmented nature of neuroscience research. 
Unlike the standardized pipelines in bioinformatics, neuroscience presents a highly heterogeneous landscape where analytical procedures and experimental datasets vary drastically across molecular, structural, functional and behavioral scales.
Analyzing such complex data requires deep domain knowledge to design appropriate analytical workflows and interpret findings  across different scales.
Crucially, the expertise necessary to ensure analytical validity~\cite{ding2025scitoolagent, wang2025geneagent} is not centralized in unified software, but instead scattered across disparate publications and isolated code repositories. 
Without an architecture that explicitly embeds this distributed domain knowledge into research workflows, general-purpose agents remain prone to methodological hallucinations~\cite{tang2025risks, messeri2024illusions}, preventing AI systems from reliably automating neuroscience research.

Here, we introduce SeekBrain, an autonomous multi-agent framework designed to accelerate neuroscience discovery through domain-grounded research planning and cross-modal data analysis. To scale domain expertise across heterogeneous data modalities, SeekBrain dynamically constructs a \textbf{Neuroscience Analysis Repertoire}. Instead of relying on static toolkits, 
the framework crystallizes methodological knowledge on demand from relevant literature and open-source codebases, encoding it into adaptable analysis recipes ranging from preprocessing to advanced modeling. 
A dual-engine architecture, comprising a Research Planning Engine and an Analysis Execution Engine, couples this structured recipe with generalized LLM reasoning to produce hierarchical research plans and orchestrate analytical pipelines across diverse neuroscience tasks. 
A researcher-in-the-loop paradigm enables experts to provide feedback that could be distilled into refined recipes and reintegrated into the repertoire, allowing the system to generalize to previously unseen challenges.

We evaluated SeekBrain using BrainArena, an expert-annotated benchmark comprising diverse data analysis tasks across species and modalities. Our framework outperformed the state-of-the-art agent frameworks Claude Code and Codex by 11.7\% and 17.7\%, respectively. Furthermore, we demonstrated the system's capacity for scientific discovery in two real-world research scenarios. In researcher-steering mode, it analyzed an multimodal dataset from free-moving zebrafish, revealing low-dimensional neural representations of behavior with structured manifold geometry and distributed brain-wide tuning. In a SeekBrain-led rediscovery task using a mouse decision-making dataset~\cite{ibl2025brainwide}, it formulated and tested a region-level decoding hypothesis, identifying a shared low-rank axis of regional decoding strength across five decoder maps, with the stimulus, choice, and feedback maps retaining spatial variation after regressing out speed and velocity. Together, these results show that SeekBrain can propose hypotheses, perform cross-modal analyses, and facilitate neuroscience discovery, providing a promising path toward AI systems that accumulate and apply domain expertise to uncover the neural principles underlying intelligence.

\section{Results}
\label{sec:results}

\begin{figure}[htbp]
    \centering
    \includegraphics[width=1.0\textwidth]{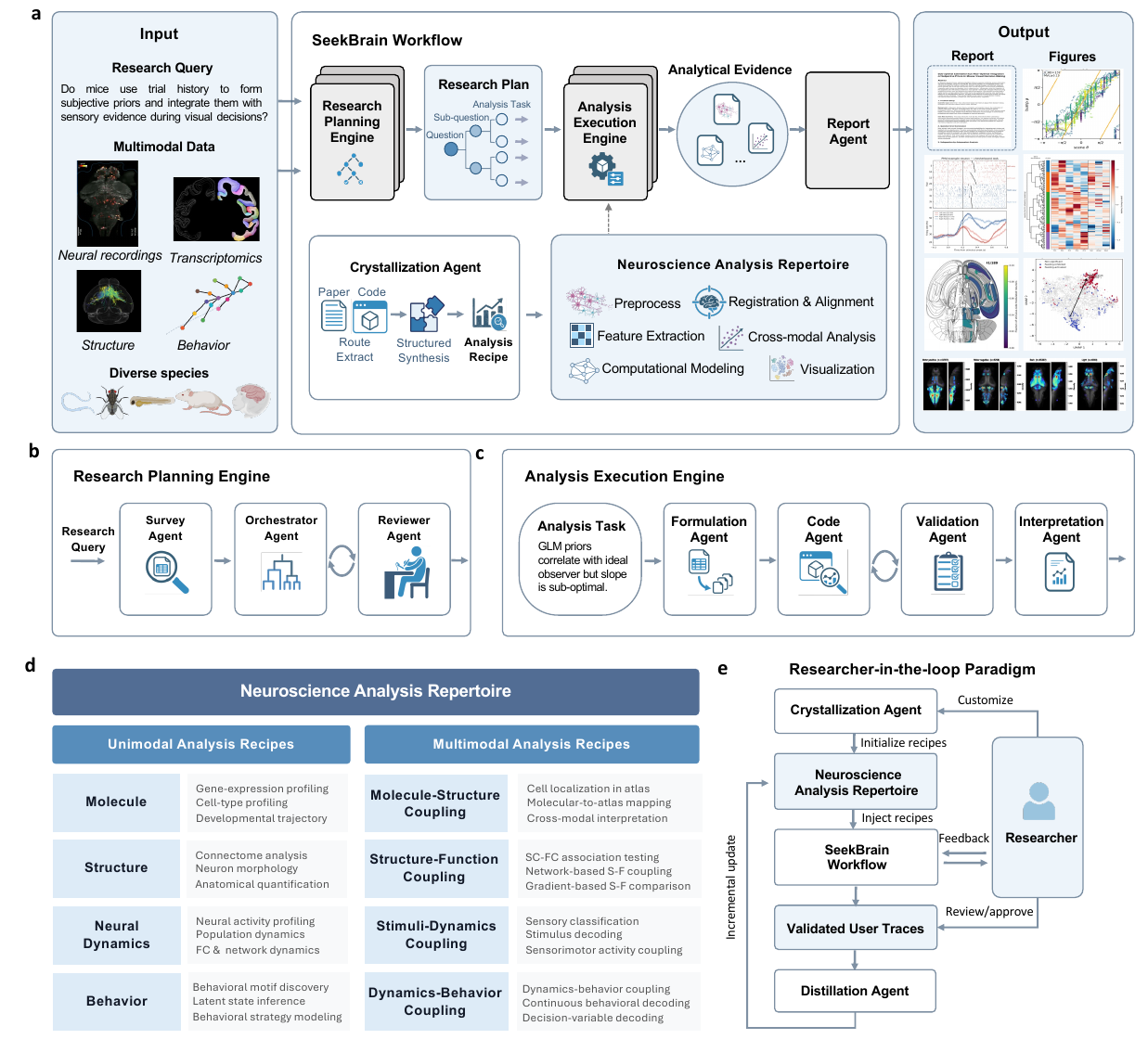}
\caption{\textbf{Overview of SeekBrain.} \textbf{a}, Overall workflow. 
SeekBrain processes research query along with neuroscience data through two core engines.
The Research Planning Engine generates a tree-structured hierarchical research plan, while the Analysis Execution Engine performs the planned analysis tasks and produces figures and statistical evidence, which are synthesized into a report. Methodological rigor is enforced by the Neuroscience Analysis Repertoire, which contains a wide range of domain-specific analysis recipes crystallized from published studies and their associated codebase. 
\textbf{b}, Research Planning Engine. After receiving a broad research query, the engine conducts an iterative debate between the Orchestrator and the Reviewer Agents to decompose the query into smaller tasks. This process yields a tree-structured research plan where terminal nodes represent specific analysis tasks for downstream execution.
\textbf{c}, Analysis Execution Engine. A Formulation Agent converts each analysis query into a structured computational schema, which the Code Agent then translates into Python scripts. A Validation Agent evaluates the outputs across four core dimensions. This closed-loop refinement process continues until the results pass validation.
\textbf{d}, Taxonomy of the Neuroscience Analysis Repertoire, which includes both unimodal and multimodal analysis recipes. Representative recipe types are presented.
FC, functional connectivity; SC, structural connectivity.
\textbf{e}, SeekBrain's researcher-in-the-loop paradigm, which enables users to flexibly guide the analysis process and supports the continuous refinement of analysis recipes based on expert interaction traces.
    }
    \label{fig:framework}
\end{figure}

\subsection{SeekBrain methodology}

The SeekBrain is a domain-specific multi-agent system that accelerates neuroscience discovery by formulating
research plans and performing cross-modal data analyses in neuroscience (Fig.~\ref{fig:framework}). To facilitate comprehensive scientific research, this framework supports a broad spectrum of multiple data modalities,
including behavior, neural activity, structural connectivity, and transcriptomic/genomic profiles, which enables diverse needs in cross-modal analytical exploration (Fig.~\ref{fig:framework}a, left). 
To ensure methodological rigor, SeekBrain is anchored by a \textbf{Neuroscience Analysis Repertoire}, which contains analysis recipes derived from published studies and refined through expert input. This repertoire crystallizes domain knowledge into reusable analysis recipes that guide automated data analysis (Fig.~\ref{fig:framework}a, middle). Using these recipes, SeekBrain operates through two agent-based engines: a \textbf{Research Planning Engine}, which breaks each general research query into specific analysis tasks organized in a hierarchical plan (Fig.~\ref{fig:framework}b), and an \textbf{Analysis Execution Engine}, which orchestrates analytical workflows, producing empirical results and scientific findings through iterative cycles of code generation, execution, and validation (Fig.~\ref{fig:framework}c). Researchers interact with SeekBrain through a natural-language interface and can guide the analysis process flexibly, establishing a collaborative researcher-in-the-loop paradigm (Fig.~\ref{fig:framework}e).

\paragraph{Neuroscience analysis repertoire.}
Data analysis in neuroscience requires specialized knowledge of data processing, modeling, statistics, and visualization. However, this knowledge is often dispersed across research papers, code repositories, and expert analysis traces rather than organized into standardized, reusable workflows. This makes it difficult for general-purpose agents to develop reliable analysis procedures from task instructions alone. 
To combine information from these sources, we developed an LLM-driven crystallization agent. The agent automatically creates analytical recipes by extracting the scientific context from research papers and the implementation details from their associated code repositories. 
Each recipe is a reusable procedure that links a scientific aim to the required data, processing and modeling steps, visualization methods, and corresponding code. Together, these recipes form the Neuroscience Analysis Repertoire, allowing SeekBrain to automatically retrieve, adapt, and combine them for new research questions. 
The initial repertoire contains 68 foundational recipes covering analysis stages from data preprocessing to visualization (Supplementary Table~\ref{tab:neuroscience_analysis_recipes}). These recipes are organized by data modality into unimodal recipes for neural dynamics, behavior, structure, and molecular data, and multimodal recipes that combine two or more modalities (Fig.~\ref{fig:framework}d). 
The repertoire is also designed to improve through use. After an adapted workflow has been reviewed and validated, its execution trace can used to update the corresponding recipe or create a new variant. This co-evolving process allows the repertoire to retain useful expert decisions and accumulate analysis experience over time.

\paragraph{Research planning engine.}
To handle high-level requests, our hierarchical planning engine performs scientific problem decomposition through a debate between an Orchestrator and a Reviewer agent (Fig.~\ref{fig:framework}b). 
With scientific background provided by a Survey Agent, the Orchestrator breaks the initial query down into targeted sub-questions (working hypotheses), which are further resolved into atomic analysis tasks.
This creates a tree-structured plan: the root node represents the original query, while terminal nodes represent specific tasks mapped to distinct biological hypotheses. This structure makes the reasoning process transparent by linking evidence from each branch directly to the main question.
To ensure scientific rigor, the Reviewer assesses the plan’s relevance to the research questions, logical coherence, data adequacy, and redundancy across tasks. 
The Orchestrator then refines the plan based on this feedback. 
This review cycle continues until both agents agree on a concrete research plan that connects the biological question to downstream analyses.

\paragraph{Analysis execution engine.}
To autonomously execute each atomic analysis task in the research plan, the Analysis Execution Engine orchestrates a four-phase workflow: formulation, coding, validation, and interpretation (Fig.~\ref{fig:framework}c).
First, a Formulation Agent evaluates the assigned task alongside the available data to generate a methodological schema. This schema details the necessary processing steps, algorithms, and visualization strategies. By retrieving and adapting recipes from the Neuroscience Analysis Repertoire, the agent aligns its approach with established domain knowledge, minimizing unsupported methodological choices.
Next, a Code Agent translates this schema into Python scripts. Through an iterative cycle of data exploration, parameter tuning, and error resolution, it generates figures and statistics to produce a reliable analytical workflow.
To prevent methodological hallucinations, a Validation Agent acts as an internal reviewer. It assesses the outputs for inquiry alignment, methodological fidelity, quantitative integrity, and visual quality. This feedback loop continues until the results are scientifically sound.
Finally, an Interpretation Agent analyzes the validated figures and statistics to formulate a self-contained interpretation summary. Throughout this process, the system transparently reports any limitations, explicitly documenting uncertainty whenever the evidence or diagnostics prove insufficient.

\begin{figure}[htbp]
    \centering
    \includegraphics[width=1.0\textwidth]{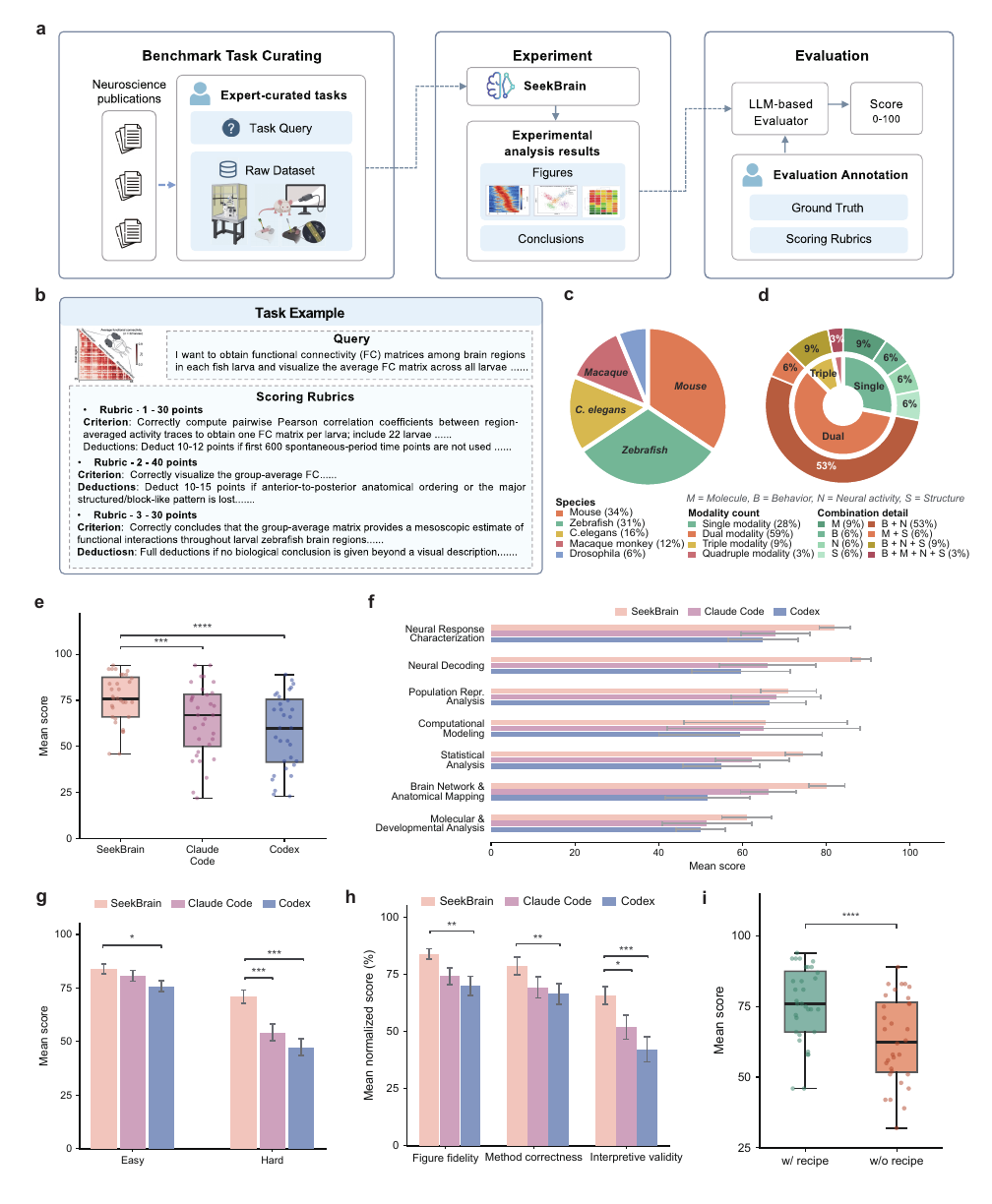}
    \caption{\textbf{BrainArena evaluation.} \textbf{a,} Overview of BrainArena construction and evaluation process. Expert-curated tasks are generated from neuroscience publications and evaluated with ground-truth annotations and scoring rubrics.
    \textbf{b,} A task example, including the task query and evaluation scoring rubrics annotated by domain experts. 
    \textbf{c,} Tasks in BrainArena span diverse model organisms in neuroscience. 
    \textbf{d,} BrainArena includes single- and multi-modality analysis tasks across molecule, behavior, neural activity, and structure modalities.}
    \label{benchmark}
\end{figure}

\begin{figure}                                                  
      \ContinuedFloat          
      \caption{\textbf{e,} Performance (mean scores) of SeekBrain and other LLM agents on BrainArena. Points indicate scores on individual tasks. Differences among the three agents were assessed using a Friedman test. Following a significant result (\(P<0.05\)), post hoc Dunn’s multiple-comparisons test was used to compare SeekBrain with Claude Code and Codex. Error bars indicate s.e.m.  
      \textbf{g,} Mean scores grouped by task difficulty: easy-level (12 tasks) and hard-level (20 tasks). Within each difficulty level, statistical analyses were performed as described for panel e.
    \textbf{h,} Component-level performance across rubric dimensions. Mean normalized scores (\% of maximum rubric points) for each rubric component are presented. The three components assess figure fidelity (visual pattern consistency with the reference), method correctness, and interpretive validity (scientific soundness of result interpretation). Within each rubric component, statistical analyses were performed as described for panel e.  
    \textbf{i,} Performance comparison between with and without recipes from neuroscience analysis repertoire. Wilcoxon signed-rank test. Statistical significance: *\(P<0.05\); **\(P<0.01\); ***\(P<0.001\), ****\(P<0.0001\).
}          
  \end{figure}

\paragraph{Researcher-in-the-loop paradigm.}
SeekBrain implements a researcher-in-the-loop paradigm (Fig.~\ref{fig:framework}e), enabling scientists to inject feedback at every stage of the workflow. Users can refine the initial research plan, assess methodological feasibility, and fine-tune analytical pipelines. 
Crucially, SeekBrain captures these user traces after each session and distills them into updated analysis recipes. By seamlessly reintegrating them into the Neuroscience Analysis Repertoire, the system continuously evolves with the broader research community.

\begin{figure}[htbp]
    \centering
    \includegraphics[width=1.0\textwidth]{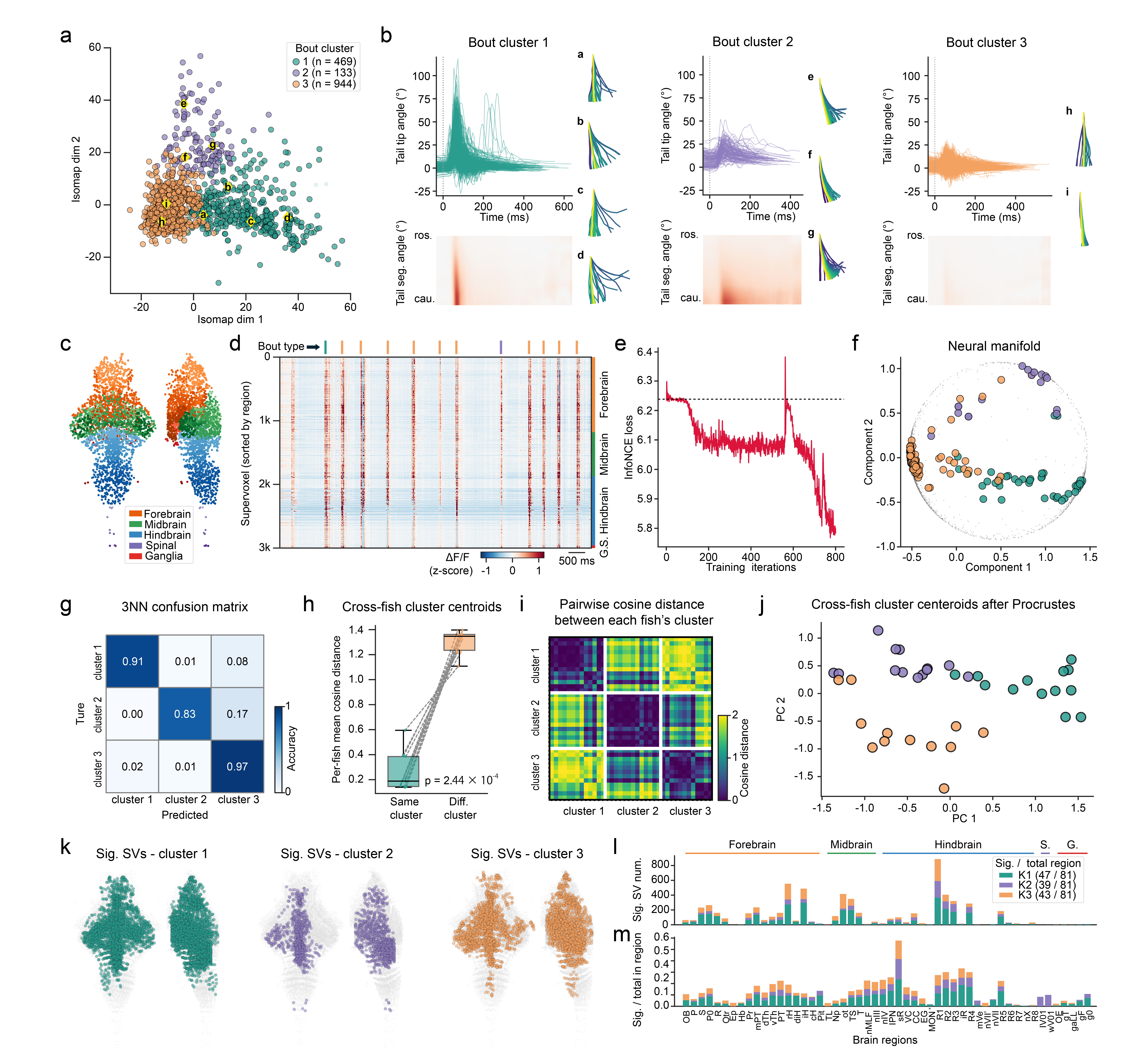}
    \caption{
    \textbf{Structured and distributed brain-wide neural representations of behavioral types in free-moving larval zebrafish.}
    \textbf{a,} Bout-level behavioral space. Each point denotes one bout (\(n = 1{,}546\) bouts from 18 larvae). Colors denote the three bout clusters. Faint points indicate long-duration bouts (>500 ms). Yellow markers a--i locate 9 representative skeletons in Panel b.
    \textbf{b,}~Tail features of 3 bout clusters. For each cluster, the upper trace plot shows all member bouts' tail-tip angle trajectories; the kinematic heatmap displays the mean time-varying angles of 9 tail segments from rostral (top) to caudal (bottom), using a fixed color scale spanning $[-90^\circ, +90^\circ]$; the right skeleton plot exhibits the representative tail posture in different bout clusters. ros., rostral; cau. caudal.
    \textbf{c,} Region-aware spatial-functional supervoxel (SV) topology from an example larva. Each point is one of \(3{,}000\) supervoxel centroids. Colors indicate major anatomical blocks. Forebrain, orange; midbrain, green; hindbrain, blue; spinal cord, purple; ganglia, red.
    \textbf{d,} Example calcium activity heatmap of brain-wide supervoxels sorted by major anatomical block and region label. Bout-locked activity peaks are denoted at the top of the heatmap. S., spinal cord; G., ganglion.
    \textbf{e,} InfoNCE training loss of the bout type supervised CEBRA model fitted in one example larva. Dashed line, uninformative-baseline reference.
    \textbf{f,}~CEBRA latent manifold for the same larva in Panel e. Gray, full-session 8D latent trajectory projected onto a 2D plotting basis derived by SVD of bout-mean latents; colored points, accepted bout-mean latents colored by bout type.
    \textbf{g,}~Aggregated 3NN confusion matrix across larvae.
    }
    \label{fig:denovo_discovery}
\end{figure}

\begin{figure}                                                             
      \ContinuedFloat          
      \caption{\textbf{h,} Cross-fish cluster centroids aligned by orthogonal Procrustes. Within each fish, the mean cosine distance to same-cluster centroids of other fish was compared to the mean cosine distance to different-cluster centroids. Boxes show fish-level summary statistics; overlaid points are fish; lines connect paired same-cluster and different-cluster values for each fish. Paired Wilcoxon signed-rank test across \(n = 13\) larvae, \(p = 2.44 \times 10^{-4}\). Diff., different.
\textbf{i,} Pairwise cosine-distance matrix among aligned fish-by-cluster centroids, grouped by cluster and fish. Color encodes pairwise cosine distance on the aligned 8D centroid vectors.
\textbf{j,} PCA projection of the same aligned centroids, colored by bout cluster type.
\textbf{k,}~Region-aware significant supervoxels for each bout type. Top and side brain projections; gray points show all supervoxel centroids pooled across larvae; colored points show supervoxels belonging to spatial clusters that were significant in the \(K\)-vs-rest contrast for each type. Sig. SVs, significant supervoxels.
\textbf{l, m} Per-region significant-supervoxel counts and ratios across larvae. The stacked display is visual only. Ratios for clusters are computed independently, not mutually exclusive. See Supplementary Table~\ref{tab:brain_region_abbr} for the full names of the brain region abbreviations.
}          
  \end{figure} 

\subsection{Assessing SeekBrain's analysis capability with BrainArena}

\paragraph{Introducing BrainArena.}
Autonomous discovery in neuroscience requires LLM agents to independently execute end-to-end data analysis. To evaluate this capability, we introduce BrainArena, a benchmark designed to test how well these agents handle complex analyses of multimodal neuroscience data (Fig.~\ref{benchmark}a).
BrainArena includes 32 tasks constructed from selected neuroscience studies (Supplementary Table~\ref{tab:brainarena_source_publications}). Each task includes an expert-annotated analysis query and a scoring rubric (Fig.~\ref{benchmark}b). The query sets the scope of the task by specifying the input data, overall objective, and required output. Each rubric is based on the original study and includes three to five scoring items, with the total score set at 100 points. The rubric evaluates three aspects: method correctness, figure fidelity (whether the visual patterns match the reference), and interpretive validity (whether the results are interpreted in a scientifically valid way). Across the 32 tasks, the rubrics contain over 700 deduction subitems that map observable errors to point deductions.

BrainArena includes data from five common model organisms (Fig.~\ref{benchmark}c): \textit{C.~elegans} (16\%), \textit{Drosophila} (6\%), zebrafish (31\%), mouse (34\%), and macaque monkey(12\%). It covers diverse data modalities, including neural activity, behavioral, anatomical and structural, and molecular and genetic data. Most tasks integrate multiple modalities: 28\% use a single modality, whereas 59\%, 9\%, and 3\% combine two, three, and four modalities, respectively; behavioral and neural activity data constitute the most common combination (53\%; Fig.~\ref{benchmark}d). These tasks span seven types of analysis: neural response characterization, neural decoding, population representation analysis, computational modeling, statistical analysis, brain network and anatomical mapping, and molecular and developmental analysis. Based on task complexity, the tasks were classified as easy or hard. Together, BrainArena combines broad coverage of common model organisms, data modalities, analysis types, and difficulty levels with expert-annotated rubrics, making it a rigorous domain-specific benchmark for neuroscience data analysis.

\paragraph{Benchmark results.}

We evaluated SeekBrain against two leading agent systems on BrainArena: Claude Code (Claude Opus 4.7~\cite{claude-opus-47}) and Codex (GPT-5.5~\cite{gpt-55}). SeekBrain uses Claude Opus 4.7 as base LLMs for most agents (details in Section~\ref{method}). The Research Planning Engine is ablated in this experiment.
Certain AI scientist frameworks are excluded from this comparison because of architectural incompatibilities: Robin~\cite{robin} and AI Scientist~\cite{aiscientist} do not support native-format scientific data, whereas AI Co-scientist~\cite{ai-coscientist} focuses on theoretical ideation and hypothesis generation.
Under these settings, SeekBrain achieved the highest mean score (75.8), compared with 64.1 for Claude Code and 58.1 for Codex; performance differed significantly among the three agents (Friedman test, \(P<0.0001\)), and post hoc Dunn’s multiple-comparisons test showed that SeekBrain outperformed both Claude Code (adjusted \(P=0.0002\)) and Codex (adjusted \(P<0.0001\); Fig.~\(\ref{benchmark}\)e).
SeekBrain also achieved the highest mean score in all seven analysis types (Fig.~\ref{benchmark}f and Supplementary Figure~\ref{benchmark-cases}). Its largest gains over Codex were in neural decoding (28.7 points) and brain network and anatomical mapping (28.5 points).
When the tasks were grouped by difficulty, SeekBrain's advantage was greater on hard tasks than on easy tasks (Fig.~\ref{benchmark}g). Performance declined more on hard tasks for the two baseline agents, whereas SeekBrain remained more consistent across the two difficulty levels.
Score decomposition showed that SeekBrain outperformed both baselines across all three evaluation dimensions (Fig.~\ref{benchmark}h). Its largest advantage was in interpretive validity, which requires accurate statistical inference and biologically grounded interpretation beyond code execution.
To further validate the role of the analysis recipes, we removed them from the workflow in an ablation study. This ablation led to issues in key analysis steps (Supplementary Table~\ref{tab:repertoire_ablation_details}) and reduced the mean score from 75.8 to 63.2 (Wilcoxon signed-rank test, $P < 0.0001$; Fig.~\ref{benchmark}i). For example, in a network visualization task, the agent failed to correctly align the spatial axes of brain region projections with those of the network node graph (Supplementary Figure~\ref{skill-cases}). Overall, these results suggest that SeekBrain's domain-grounded engine automatically performs diverse,
end-to-end analyses, consistently producing faithful figures and scientifically meaningful interpretations.

\begin{figure}[t]
    \centering
    \includegraphics[width=1.0\textwidth]{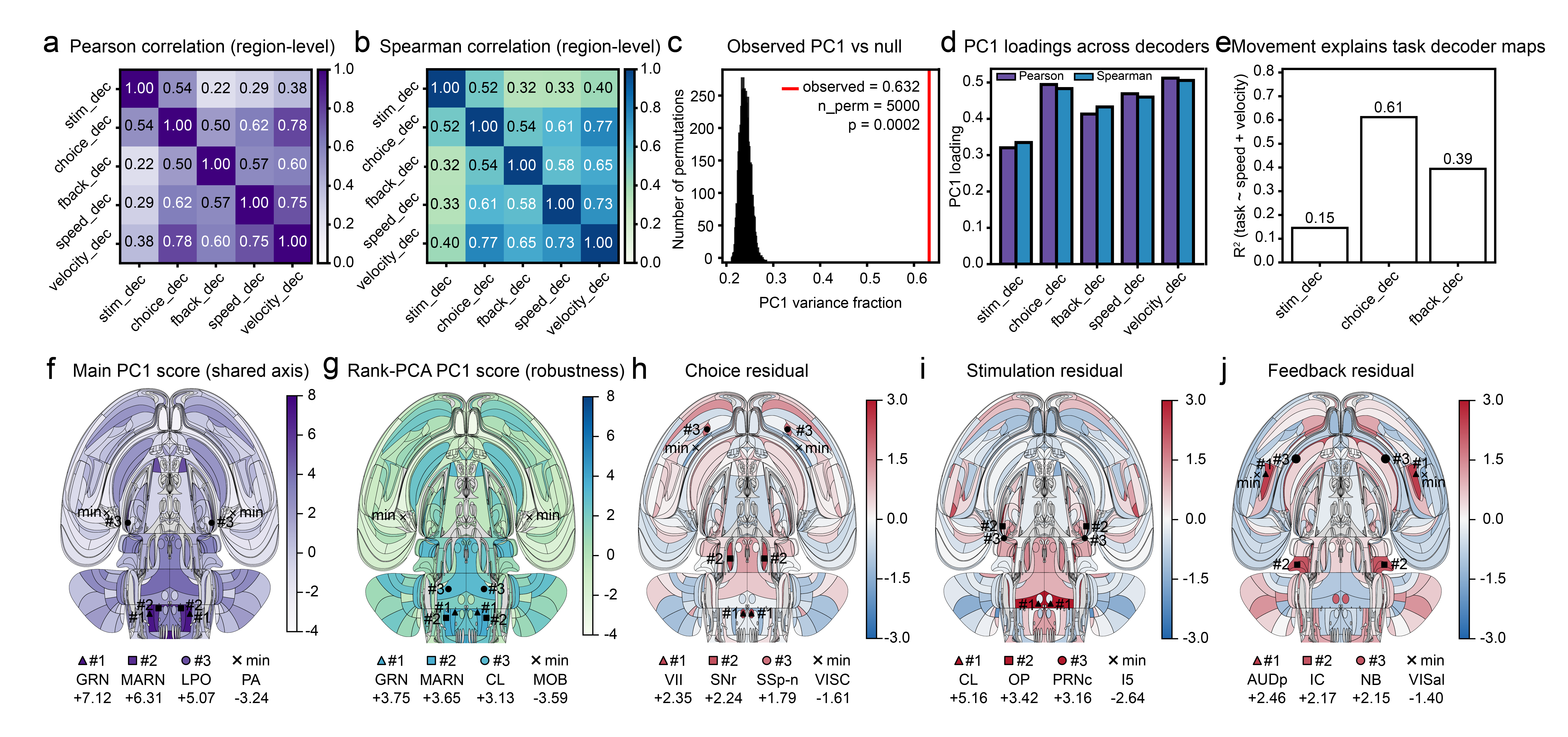}
    \caption{
    \textbf{A shared axis of regional decoding strength and task-specific residual patterns in a mouse decision-making task.}
    \textbf{a,} Pearson correlation matrix of stimulus, choice, feedback, speed, and velocity decoder maps across 201 regions.
    \textbf{b,} Spearman rank-correlation matrix for the same five decoders.
    \textbf{c,} Observed PC1 variance fraction (red line; 0.632) compared with a 5,000-permutation column-shuffle null; empirical \(P \approx 0.0002\).
    \textbf{d,} PC1 loadings for the metric-value PCA and rank-PCA decompositions; all decoder loadings are positive.
    \textbf{e,} \(R^{2}\) from ordinary least-squares regression of each task decoder on speed and velocity.
    \textbf{f,} Main PC1 scores for the 201 included regions, rendered bilaterally and mirror-symmetrically on the Swanson mouse-brain flatmap. Higher scores occurred in reticular, motor-related brainstem, and thalamic regions.
    \textbf{g,} Rank-PCA PC1 score projected onto the same atlas, confirming robustness of the shared axis.
    \textbf{h,} Choice-decoder residual map after removing speed- and velocity-linked variance.
    \textbf{i,} Stimulus-decoder residual map after removing speed- and velocity-linked variance.
    \textbf{j,} Feedback-decoder residual map after removing speed- and velocity-linked variance. See Supplementary Table~\ref{tab:brain_region_abbr} for the full names of the brain region abbreviations.
    }
    \label{fig:ibl_decoder_low_rank_axis}
\end{figure}

\subsection{Case study 1: SeekBrain analyzes a tri-modal zebrafish dataset through a sequence of researcher-led questions}

To show how SeekBrain supports complex scientific investigations that require multiple linked stages of analysis, we applied it to a multimodal dataset capturing brain-wide neuronal calcium dynamics and free-swimming behavior in larval zebrafish. We used SeekBrain to perform integrated analyses linking behavior, neural activity, and anatomy. At each stage, the researcher posed a question based on earlier results, and the agent ran the corresponding analysis pipeline.

First, to determine whether swimming bouts could be mapped into a behavioral space based on tail kinematics, the researcher asked the agent to run an adapted behavioral clustering pipeline based on an established framework \cite{mearns2020deconstructing}. The agent generated and ran all scripts required for this multistep analysis. The analysis included decomposing time-varying tail-segment angles into principal components (PCs) of posture, calculating polarity-invariant dynamic time warping (DTW) distances, embedding the resulting distance matrix with Isomap, and identifying bout types with Ward’s hierarchical clustering. Consistent with previous studies, larval swimming bouts occupied a continuous behavioral space (Fig.~\ref{fig:denovo_discovery}a). Across the tested values of \(K\), the silhouette score was highest at \(K=3\), supporting a three-cluster partition (Fig.~\ref{fig:denovo_discovery}a,b). Cluster 1 contained large-angle turns, Cluster 2 corresponded to J-turns, and Cluster 3 consisted mainly of low-amplitude forward swimming (Fig.~\ref{fig:denovo_discovery}b).

To link high-dimensional, brain-wide neural dynamics to these behavioral clusters, we next sought to identify a latent neural manifold and isolate candidate brain regions associated with each cluster. The raw dataset contained 64,032 $\pm$ 8,575 ROIs per fish (mean $\pm$ s.d.). To reduce the data size while retaining spatial and functional information, the agent grouped ROIs across the brain into region-aware supervoxels (SVs). Within each anatomical region, k-means clustering was applied to a feature matrix that combined PCs of neural activity with the spatial coordinates of the ROIs. This procedure produced a fixed total of 3,000 SVs per larva (Fig.~\ref{fig:denovo_discovery}c). The activity traces from these region-aware SVs were then aligned to bout epochs for subsequent cross-modal analyses (Fig.~\ref{fig:denovo_discovery}d).

To test whether a low-dimensional representation representation of brain-wide activity could capture the three behavioral clusters, the researcher asked the agent to train a CEBRA model supervised by bout type \cite{schneider2023learnable} using behaviorally relevant frames. The trained model was then applied to the full time series from each fish (Fig.~\ref{fig:denovo_discovery}e,f). For every fish, training converged to an InfoNCE loss below the uninformative baseline; the final loss was $5.703 \pm 0.070$ ($\text{mean} \pm \text{s.d.}$). When averaged within bouts, the resulting eight-dimensional (8D) neural representations formed clearly separated clusters (Fig.~\ref{fig:denovo_discovery}f). In cross-validation, a 3-nearest-neighbor (3NN) classifier decoded the behavioral cluster labels from the CEBRA representations with an accuracy of $0.949 \pm 0.034$ ($\text{mean} \pm \text{s.d.}$; Fig.~\ref{fig:denovo_discovery}g). These results show that the learned neural representation captured behavior-related differences within individual fish.

To compare the latent cluster geometry across fish, we asked the agent to align the cluster centroids using orthogonal Procrustes analysis. Only clusters with at least three bouts were included. After alignment, the agent calculated, for each fish, the mean cosine distances between its cluster centroids and matching or nonmatching centroids from the other fish. The mean distance between matching centroids was significantly smaller than that between nonmatching centroids (paired Wilcoxon signed-rank test, \(p = 2.44 \times 10^{-4}\), \(n = 13\) larvae; Fig.~\ref{fig:denovo_discovery}h). This shared cluster organization across fish was also visible in the pairwise cosine-distance matrix and the PC projection of the aligned centroids (Fig.~\ref{fig:denovo_discovery}i,j).

Finally, to map the spatial distribution of neural activity associated with each behavioral cluster, we asked the agent to run a region-aware permutation test to identify SVs whose bout-averaged calcium activity differed significantly between that cluster and the other two clusters combined (Fig.~\ref{fig:denovo_discovery}k). These SVs were distributed throughout the brain, from the forebrain to the hindbrain. They were found in 47, 39, and 43 of the 81 anatomical regions for Clusters 1, 2, and 3, respectively (Fig.~\ref{fig:denovo_discovery}l,m). Some regions contained a relatively high proportion of tuned SVs, including the thalamus, pretectum (PT), several hypothalamic subregions, the nucleus of the medial longitudinal fasciculus (nMLF), the superior and inferior raphe nuclei, and several hindbrain rhombomeres (Fig.~\ref{fig:denovo_discovery}m).

Taken together, this case study discovered the structured and distributed brain-wide neural representations of behavioral types in free-moving larval zebrafish. Throughout this data exploration, SeekBrain reliably carried out the required analyses, helping the researcher investigate the neural basis of behavior.

\subsection{Case study 2: SeekBrain reanalyzes a mouse decision-making dataset through agent-led hypothesis testing}

While the first case study focused on SeekBrain’s ability to carry out a sequence of analyses defined by the researcher, the second tested whether it could propose testable hypotheses and plan analyses to evaluate them. We therefore used SeekBrain to reanalyze the published International Brain Laboratory (IBL) mouse decision-making dataset \cite{ibl2025brainwide} in an agent-led process supervised by the researcher. SeekBrain proposed candidate hypotheses and analysis plans, while the researcher selected promising directions and gave specific guidance on implementation.

Findling et al. used this dataset to examine how task priors are represented in the brain during visual decision-making \cite{findling2025prior}. The mice inferred the changing probability of a stimulus appearing on either side and used this information to guide their choices. This internal estimate, termed the subjective prior, was encoded in 20--30\% of brain regions. The widespread representation of this prior was consistent with distributed recurrent Bayesian inference, rather than with prior integration limited to decision-making regions.

SeekBrain proposed another region-level hypothesis using the same dataset: decoder performance for stimulus, choice, feedback, speed, and velocity might share a dominant pattern across brain regions rather than vary independently. To test this hypothesis, the agent used data from 279 unique Beryl-labeled regions in the original region-level table from Findling et al. \cite{findling2025prior}. It retained the 201 regions with finite values for all five decoder variables, forming a \(201 \times 5\) region-by-decoder matrix.

Pairwise correlations first suggested that the five decoder maps shared a common pattern across brain regions. Both Pearson and Spearman analyses showed generally positive correlations: regions with high decoding strength for one variable also tended to have high decoding strength for the others (Fig.~\ref{fig:ibl_decoder_low_rank_axis}a,b). SeekBrain then standardized each decoder column using z-scores and applied principal component analysis (PCA) to the resulting region-by-decoder matrix. The PC1 explained 63.2\% of the total variance. This value was higher than expected under a null distribution generated by independently shuffling the values within each decoder column 5,000 times (empirical $P \approx 2.0 \times 10^{-4}$; Fig.~\ref{fig:ibl_decoder_low_rank_axis}c). The loadings of all five decoder variables on PC1 had the same sign, supporting a shared axis of decoding strength rather than an opposing pattern among the variables (Fig.~\ref{fig:ibl_decoder_low_rank_axis}d).

To assess whether the shared-axis pattern depended on the scale of the decoder values or on extreme values, SeekBrain repeated the PCA after replacing the values in each decoder map with their ranks within that map. After matching the orientation of PC1 across the two analyses, the regional PC1 scores from rank-based PCA were strongly correlated with those from the original value-based PCA ($r = 0.96$). The two analyses also produced similar spatial patterns of regional PC1 scores in the brain atlas (Fig.~\ref{fig:ibl_decoder_low_rank_axis}f,g), showing that the shared axis was robust to differences in scale and to extreme values. High shared-axis scores were found in reticular regions, motor-related brainstem areas, and thalamic regions, whereas several olfactory and piriform-amygdalar regions had lower scores.

SeekBrain next asked whether this shared axis reflected the widespread influence of movement on neural activity. It used ordinary least-squares regression to predict each task decoder map from the speed and velocity decoder strengths. These movement variables explained 61\% of the regional variance in choice decoding (\(R^2 = 0.61\)), 39\% in feedback decoding (\(R^2 = 0.39\)), and 15\% in stimulus decoding (\(R^2 = 0.15\); Fig.~\ref{fig:ibl_decoder_low_rank_axis}e). After accounting for movement, the residual maps still showed distinct regional patterns for choice, stimulus, and feedback (Fig.~\ref{fig:ibl_decoder_low_rank_axis}h-j). Thus, although movement was associated with the shared decoding pattern, it did not fully explain the regional patterns specific to each task variable.

In summary, this case study demonstrates that SeekBrain can propose and test new hypotheses using published datasets, helping researchers explore existing data in new ways.

\section{Discussion}
\label{discussion}

In this work, we developed SeekBrain, a specialized multi-agent framework that accelerates discovery in neuroscience through domain-grounded research planning and cross-modal data analysis. Crucially, our approach overcomes a fundamental bottleneck in deploying AI for neuroscience: the severe mismatch between generic agent architectures and the fragmented nature of neuroscientific research. To bridge this gap, we constructed a dynamic Neuroscience Analysis Repertoire that autonomously crystallizes dispersed expertise into analysis workflows. Driven by hierarchical planning and analysis execution engines, this architecture successfully transforms general-purpose LLMs into scientific collaborators capable of proposing plausible exploratory plans and orchestrating analysis workflow. We validated this capability through comprehensive benchmarking and two real-world cross-modal discovery cases, highlighting its potential to automate complex research pipelines.

SeekBrain diverges from existing LLM-based scientific agents by prioritizing the internalization of domain-specific methodology. Whereas frameworks like AI Co-Scientist~\cite{ai-coscientist} or AI Scientist~\cite{aiscientist} rely on generic reasoning without grounding execution in field-specific constraints, and bioinformatics agents like CellVoyager~\cite{cellvoyager} inherit established software abstractions (e.g., scanpy~\cite{scanpy}), our system must construct its expertise largely from scratch. By distilling analysis recipes directly from literature and open-source codebases, our framework creates an indexable, evolvable knowledge base. Future iterations will focus on broadening this repertoire to include spatial transcriptomics, large-scale connectomics, and closed-loop behavior. 

The BrainArena, as a domain-specific benchmark to evaluate automatic analysis ability on real-world data, complements general academic-reasoning benchmarks such as Humanity's Last Exam~\cite{hle} and code-centric benchmarks such as SWE-bench~\cite{swebench} and MLE-bench~\cite{mlebench}. Its 32 expert-annotated tasks are scored using rubrics comprising more than 700 deduction subitems across figure fidelity, method correctness, and interpretive validity, making BrainArena a rigorous benchmark for assessing agents against the demands of end-to-end scientific workflows. 
On this benchmark, SeekBrain outperformed Claude Code and Codex across all seven analytical domains, with its advantage becoming more pronounced on harder tasks. Among the three scoring dimensions, its largest advantage was in interpretive validity, which assesses whether agents can draw scientifically valid and meaningful conclusions from their analytical results. 
Ablating the repertoire from the framework caused a significant performance drop of over 10 points, demonstrating that agents struggle to derive valid biological interpretations without explicit domain knowledge to enforce scientific rigor.

Beyond the quantitative benchmark, two real-world case studies demonstrate SeekBrain's capability to drive scientific exploration in different modes.
Under researcher guidance, SeekBrain analyzed the multimodal dataset from free-moving larval zebrafish. Building on Mearns et al.’s study of swimming behavior ~\cite{mearns2020deconstructing}, it examined how different types of swimming bouts were represented across the brain. Through a sequence of instructed analyses, SeekBrain revealed a low-dimensional structure linking neural activity and behavior.
In a more autonomous setting, SeekBrain reanalyzed decoder outputs~\cite{findling2025prior} from the IBL mouse dataset~\cite{ibl2025brainwide} and tested whether decoding strength shared a common pattern across brain regions. Previous studies had shown that movement broadly affects neural dynamics~\cite{stringer2019spontaneous, musall2019singletrial, steinmetz2019distributed}. On this background, SeekBrain represented five published region-wise decoder maps (stimulus, choice, feedback, speed, and velocity) along a shared low-rank axis of decoding strength. It then tested whether this shared axis was related to movement and separated the movement-related component from the remaining task-specific regional patterns.  
Together, these cases demonstrate that SeekBrain supports real-world research across varying levels of automation, enabling researchers to maintain control over key scientific decisions while delegating routine analysis and exploration.
More broadly, large-scale brain initiatives and data-sharing efforts of the community have created an expanding ecosystem of open-access, multimodal datasets~\cite{biccn2021multimodal, yao2023highresolution, microns2025functional, dorkenwald2024neuronal, ibl2025brainwide,rubel2022nwb}, yet much of which remains underexplored.
Equipped with dynamic domain-specific repertoire, 
SeekBrain serves as a evolving system to generate new hypotheses and systematically reanalyze these public datasets, ultimately helping uncover biological principles hidden within the growing body of neuroscience data resources.

Looking ahead, SeekBrain points toward a continuous development loop between AI-driven neuroscience and biologically inspired AI. Within neuroscience, the framework provides a foundation for a scalable, collaborative ecosystem where research groups can seamlessly contribute and curate specialized analytical workflows. Conversely, on the AI development side, extracting the computational principles underlying biological intelligence offers vital inductive biases for next-generation artificial models. We envision that this feedback loop will create a future where intelligent systems accelerate our understanding of the brain, and in turn, use those insights to guide their own evolution.

\section{Methods}
\label{method}

\subsection{Research planning engine}
\label{sec4.1}

We address this limitation by introducing a hierarchical planning workflow that explicitly models this progressive decomposition. Every stage of the process incorporates an automated peer-review mechanism that evaluates scientific validity, data adequacy, and execution feasibility, providing iterative feedback to refine the plan. We use Gemini-3.1-Pro~\cite{gemini-31-pro} for sub-question generation, question-level synthesis and reporting, whereas Claude-Opus-4.7~\cite{claude-opus-47} is used for review, analysis task generation and sub-question-level conclusion synthesis.

\paragraph{Research background survey.} Before research planning, a Survey Agent constructs the required scientific background from the literature corpus. Given a research query, the agent identifies relevant publications, synthesizes key findings and unresolved gaps into a concise background summary via retrieval-augmented generation. 
Literature-grounded planning produces scientifically motivated research plans that build on prior work and target meaningful open questions.
The system also supports user-provided background context.

\paragraph{Sub-question decomposition and task generation.}
The Orchestrator Agent converts a broad scientific objective into a structured research plan.
It takes three structured inputs: a user-defined scientific objective, the target dataset metadata, and optional background information.
If the metadata is absent, a Data Analysis Agent is invoked to autonomously profile the dataset, extracting data structures, variable types, sample sizes, and etc. 
The Orchestrator Agent then divides the main question into related sub-questions and creates concrete analysis tasks for each one. 
These sub-questions and tasks must be testable, biologically meaningful, non-redundant, and feasible with the available data.
The resulting plan has a tree structure in which the original query forms the root node, and the analysis tasks form terminal nodes.
Users may specify the number of sub-questions and tasks or allow the system to determine it based on task complexity.

\paragraph{Automated iterative peer-review.} To ensure the workflow's robustness, we implemented a multi-agent iterative refinement loop following both the sub-question and analysis task decomposition stages. 
An independent Reviewer Agent evaluates the Orchestrator's outputs based on three criteria: scientific validity, data adequacy, and execution feasibility. 
The Reviewer assigns a quantitative score to each criterion. If a proposed question or task falls below the acceptance threshold, the Reviewer synthesizes targeted feedback and revision suggestions. These comments are then fed back into the Orchestrator's context to guide the next round of generation. This generation-review-revision loop continues until the output meets the acceptance criteria or reaches a predefined limit (\textit{e.g.}, $N=3$ iterations).

\subsection{Analysis execution engine}
\label{sec4.2}
SeekBrain coordinates a set of specialized language agents that convert analysis task query into executable data analysis workflows. 
The coding and core reasoning agents in this engine are powered by Claude Opus 4.7~\cite{claude-opus-47} with adaptive extended thinking enabled; long-text auxiliary tasks are routed to Gemini 3.1 Pro~\cite{gemini-31-pro}.

\paragraph{Analytical formulation.}
Given an neuroscience analysis inquiry, we design a Formulation Agent to produce a step-by-step detailed methodological schema.
The agent evaluates the inquiry against the empirical data and retrieves analysis recipes from constructed neuroscience analysis repertoire. Under the guidance of analysis recipes, it decomposes the request into a structured sequential methodological schema, explicitly detailing the data inputs, expected output and execution specifications. 
Acting as a strict computational constraint, the generated schema anchors the subsequent workflow, ensuring downstream agents accurately translate it into executable scripts.


\paragraph{Closed-loop generation and validation.}
To automate the neuroscience data analysis, we design a generation and validation loop to generate robust analysis workflow scripts. In the generation phase,
the Code Agent translates the generated analysis plan into executable Python scripts. The Code Agent employs a ReAct (Reasoning and Acting) paradigm~\cite{react} to autonomously write, execute, and debug the analytical code under restricted reasoning and execution iterations. The process iterates between data probing, parameter tuning, statistical evaluation, and runtime error resolution before ultimately yielding robust scripts grounded in comprehensive data exploration. The scripts are executed to output scientific figures and key quantitative statistics.

In the validation phase, SeekBrain includes 
a Validation Agent that enforces research rigor by evaluating whether the outputs constitute a scientifically valid neuroscience analysis, which distinguishes it from general-purpose agentic systems.
Following each successful execution round, the agent assesses the generated results across four dimensions: (1) alignment with the original scientific inquiry, (2) methodological fidelity to the analysis plan, (3) quantitative data integrity, and (4) visual quality of the generated figures. 
Any failure to meet these criteria triggers an iterative refinement loop, with the Validation Agent supplying its diagnostic feedback to the Coding Agent. This generation-and-validation loop iterates across several rounds until the analytical outputs achieve full compliance, ensuring the final results are both methodologically robust and biologically meaningful.

\paragraph{Result interpretation.}
The Interpretation Agent evaluates the generated figures and quantitative statistics to extract core scientific insights. It then integrates these insights and full execution traces to formulate a comprehensive scientific summary of the entire analytical process. It explicitly reports method limitations whenever analysis fails, statistical evidence proves insufficient, or numerical diagnostics trigger warnings.

\subsection{Report generation}

Completing the analysis tasks yields a large set of intermediate outputs, including statistics, figures, tables, and text. To handle this, a Report Agent aggregates the evidence hierarchically to form traceable conclusions. It begins by summarizing each analysis task results into key findings, supporting statistics, visuals, and limitations. It then merges the evidence from related plans to answer each sub-question, keeping clear references to the original plans. Finally, the agent combines these sub-question answers to address the main scientific objective. 
Using this aggregated evidence, the agent generates a formatted academic PDF report with standard sections.

\subsection{Construction of the Neuroscience Analysis Repertoire}
To bridge the knowledge gap between SeekBrain and neuroscience expert,  
the Neuroscience Analysis Repertoire acts as a dynamic knowledge base by continually expanding through two ways:  crystallization analytical procedures from open-sourced literature, and the subsequent distillation of validated execution traces guided by user feedback.

\paragraph{Recipe crystallization.}
To construct the repertoire autonomously, a Crystallization Agent extracts and synthesizes methodological knowledge from paired inputs comprising open-access publications and their corresponding source-code repositories.
For a specified analytical objective, the agent extracts task-relevant methodological descriptions from the manuscript and simultaneously scans the repository to identify relevant executable components. 
The agent then performs a cross-modal alignment, explicitly mapping theoretical analytical steps described in the text to specific functional implementations and code blocks in the repository. 
Once this correspondence is established, the agent compiles the theoretical context and the executable code into a standardized recipe. This formatted recipe integrates the scientific rationale with specific computational operations, which is subsequently reintegrated into the repertoire for future neuroscience analysis.

\paragraph{Distillation of user trajectories.}
The repertoire could continuously evolves with its users through a Distillation Agent. 
Upon the successful completion of an analytical workflow, the system invokes a Distillation Agent to process the computational history.   
The agent receives the complete execution trace, which comprises the finalized code blocks, runtime logs, and any qualitative feedback provided by the human user. 
To derive a reusable analysis recipe, the Distillation Agent parses these inputs to abstract the core analytical logic. Specifically, the agent is prompted to retain the sequence of methodological operations and expected outputs, while stripping away incidental, run-specific artifacts such as local file paths or dataset-dependent variable names. 
The abstracted procedure is then formatted into a standardized recipe and appended to the Neuroscience Analysis Repertoire. This update mechanism allows the system to progressively expand its methodological database using validated researcher-in-the-loop trajectories.

\subsection{BrainArena}

\paragraph{Benchmark construction.}
BrainArena built on source studies with publicly available neuroscience data and reproducible figure-level analyses.
The benchmark contained 32 tasks from 10 source studies~\cite{Atanas2023BrainWide,findling2025prior,ibl2025brainwide,Tanaka2026Plastic,Basnak2025Multimodal,Gao2025Continuous,Genkin2025Dynamics,Yu2025Cholinergic,Legare2025Structural,MarquezLegorreta2026WholeBrain}, with each task corresponding to a key figure in the original publication.
Each task was annotated with an expert-written analysis query, data paths, a reference figure and a ground-truth scientific conclusion.
Domain experts checked the task specification against the source study and constructed a task-specific 100-point scoring rubric.
Supplementary Table~\ref{tab:neurosci-bench-tasks} provides the complete task metadata.

\paragraph{Experimental setup.}
For inference on BrainArena, each system received the same task package, consisting only of the analysis task query and raw data paths.
To ensure rigorous and unbiased evaluation, all tested systems were strictly blinded to the source paper identity, reference figures, and ground-truth conclusions.
We evaluated SeekBrain, Claude Code and Codex on BrainArena.
To mitigate the risk of benchmark data leakage, BrainArena included tasks derived from a bioRxiv preprint that was published after the official release date of the Claude model.
The task execution prompt used for benchmark baselines is provided in Supplementary Figure~\ref{fig:prompt_task_execution}.

\paragraph{Evaluation.}
For evaluation, we used Claude Opus 4.6~\cite{claude-opus-46} as a rubric-based LLM judge to assess whether the model outputs reproduced the reference figures and scientific conclusions.
The judge received the task query, model output, and returned item-level scores based on task-specific rubrics annotated by experts.
The LLM-as-judge evaluation prompt is provided in Supplementary Figure~\ref{fig:prompt_evaluation}.

\begingroup
\sloppy

\clearpage
\printbibliography[heading=bibintoc]

@article{biccn2021multimodal,
  title   = {A multimodal cell census and atlas of the mammalian primary motor cortex},
  author  = {{BRAIN Initiative Cell Census Network (BICCN)}},
  journal = {Nature},
  volume  = {598},
  number  = {7879},
  pages   = {86--102},
  year    = {2021}
  %doi     = {10.1038/s41586-021-03950-0}
}

@article{finn2023functional,
  title={Functional neuroimaging as a catalyst for integrated neuroscience},
  author={Finn, Emily S. and Poldrack, Russell A. and Shine, James M.},
  journal={Nature},
  volume={623},
  number={7986},
  pages={263--273},
  year={2023}
}

@article{yao2023highresolution,
  title   = {A high-resolution transcriptomic and spatial atlas of cell types in the whole mouse brain},
  author  = {Yao, Zizhen and van Velthoven, Cindy T. J. and Kunst, Michael and Zhang, Meng and McMillen, Delissa and Lee, Changkyu and Jung, Won and Goldy, Jeff and Abdelhak, Aliya and Aitken, Matthew and others},
  journal = {Nature},
  volume  = {624},
  number  = {7991},
  pages   = {317--332},
  year    = {2023}
  %doi     = {10.1038/s41586-023-06812-z}
}

@article{microns2025functional,
  title     = {Functional connectomics spanning multiple areas of mouse visual cortex},
  author    = {{The MICrONS Consortium}},
  journal   = {Nature},
  volume    = {640},
  number    = {8058},
  pages     = {435--447},
  year      = {2025}
  %doi       = {10.1038/s41586-025-08790-w}
}

@article{dorkenwald2024neuronal,
  title     = {Neuronal wiring diagram of an adult brain},
  author    = {Dorkenwald, Sven and Matsliah, Arie and Sterling, Amy R. and Schlegel, Philipp and Yu, Szi-chieh and McKellar, Claire E. and Lin, Albert and Costa, Marta and Eichler, Katharina and Yin, Yijie and others},
  journal   = {Nature},
  volume    = {634},
  number    = {8032},
  pages     = {124--138},
  year      = {2024},
  publisher = {Nature Publishing Group},
  %doi       = {10.1038/s41586-024-07558-y}
}

@article{ibl2025brainwide,
  title     = {A brain-wide map of neural activity during complex behaviour},
  author    = {{International Brain Laboratory} and Angelaki, Dora and Benson, Brandon and Benson, Julius and Birman, Daniel and Bonacchi, Niccol{\`o} and Bougrova, Kc{\'e}nia and Bruijns, Sebastian A. and Carandini, Matteo and Catarino, Joana A. and Chapuis, Gaelle A. and others},
  journal   = {Nature},
  volume    = {645},
  number    = {8079},
  pages     = {177--191},
  year      = {2025},
  publisher = {Nature Publishing Group},
  %doi       = {10.1038/s41586-025-09235-0}
}

@article{arkhipov2025integrating,
  title     = {Integrating multimodal data to understand cortical circuit architecture and function},
  author    = {Arkhipov, Anton and da Costa, Nuno and de Vries, Saskia and Bakken, Trygve and Bennett, Corbett and Bernard, Amy and Berg, Jim and Buice, Michael and Collman, Forrest and Daigle, Tanya and others},
  journal   = {Nature Neuroscience},
  volume    = {28},
  number    = {4},
  pages     = {717--730},
  year      = {2025},
  publisher = {Nature Publishing Group},
  %doi       = {10.1038/s41593-025-01904-7}
}

@article{mathis2026joint,
  title   = {Joint modelling of brain and behaviour dynamics with artificial intelligence},
  author  = {Mathis, Mackenzie Weygandt and Mathis, Alexander},
  journal = {Nature Reviews Neuroscience},
  volume  = {27},
  number  = {2},
  pages   = {87--100},
  year    = {2026},
  %doi     = {10.1038/s41583-025-00996-1}
}

@article{robin,
  title     = {A multi-agent system for automating scientific discovery},
  author    = {Ghareeb, Ali Essam and Chang, Benjamin and Mitchener, Ludovico and Yiu, Angela and Szostkiewicz, Caralyn J. and Shved, Dmytro and Gyimesi, Gavin J. and Laurent, Jon M. and Wright, Samantha M. and Razzak, Muhammed T. and White, Andrew D. and Finnemann, Silvia C. and Hinks, Michaela M. and Rodriques, Samuel G.},
  journal   = {Nature},
  volume    = {655},
  number    = {8122},
  pages     = {497--505},
  year      = {2026},
  publisher = {Nature Publishing Group}
  %doi       = {10.1038/s41586-026-10652-y}
}

@article{aiscientist,
  title={Towards end-to-end automation of AI research},
  author={Lu, Chris and Lu, Cong and Lange, Robert Tjarko and Yamada, Yutaro and Hu, Shengran and Foerster, Jakob and Ha, David and Clune, Jeff},
  journal={Nature},
  volume={651},
  number={8107},
  pages={914--919},
  year={2026},
  publisher = {Nature Publishing Group}
  %doi       = {10.1038/s41586-026-10265-5}
}

@article{ai-coscientist,
  title={Accelerating scientific discovery with Co-Scientist},
  author={Gottweis, Juraj and Weng, Wei-Hung and Daryin, Alexander and Tu, Tao and Sirkovic, Petar and Myaskovsky, Artiom and Glowaty, Grzegorz and Weissenberger, Felix and Orlandi, Alessio and Popovici, Dan and others},
  journal   = {Nature},
  volume    = {655},
  number    = {8122},
  pages     = {487--496},
  year      = {2026},
  publisher = {Nature Publishing Group}
  %doi       = {10.1038/s41586-026-10644-y}
}

@article{cellvoyager,
  title={Cellvoyager: Ai compbio agent generates new insights by autonomously analyzing biological data},
  author={Alber, Samuel and Chen, Bowen and Sun, Eric and Isakova, Alina and Wilk, Aaron J. and Zou, James},
  journal   = {Nature Methods},
  volume    = {23},
  number    = {4},
  pages     = {749--759},
  year      = {2026},
  publisher = {Nature Publishing Group}
  %doi       = {10.1038/s41592-026-03029-6}
}

@article{
huang2025biomni,
author = {Kexin Huang  and Serena Zhang  and Hanchen Wang  and Yuanhao Qu  and Yingzhou Lu  and Ryan Li  and Yusuf Roohani  and Lin Qiu  and Shiyi Cao  and Gavin Li  and Junze Zhang  and Di Yin  and Rick Wierenga  and Deniz Kavi  and Sherry Liu  and Tianwei She  and Shruti Marwaha  and Jennefer N. Carter  and Xin Zhou  and Matthew T. Wheeler  and Jonathan A. Bernstein  and Mengdi Wang  and Peng He  and Jingtian Zhou  and Michael P. Snyder  and Le Cong  and Aviv Regev  and Jure Leskovec },
title = {Autonomous biomedical research with an artificial intelligence agent},
journal = {Science},
pages = {eadz4351},
year = {2026},
doi = {10.1126/science.adz4351}
%eprint = {https://www.science.org/doi/pdf/10.1126/science.adz4351}
}

@article{software,
  title={An AI system to help scientists write expert-level empirical software},
  author={Ayg{\"u}n, Eser and Belyaeva, Anastasiya and Comanici, Gheorghe and Coram, Marc and Cui, Hao and Garrison, Jake and Johnston, Renee and Kast, Anton and McLean, Cory Y and Norgaard, Peter and others},
  journal   = {Nature},
  volume    = {654},
  number    = {8120},
  pages     = {909--916},
  year      = {2026},
  publisher = {Nature Publishing Group}
  %doi       = {10.1038/s41586-026-10658-6}
}

@inproceedings{react,
  title={ReAct: Synergizing Reasoning and Acting in Language Models},
  author={Yao, Shunyu and Zhao, Jeffrey and Yu, Dian and Du, Nan and Shafran, Izhak and Narasimhan, Karthik and Cao, Yuan},
  booktitle={International Conference on Learning Representations},
  year={2023}
}

@article{scanpy,
  title={SCANPY: large-scale single-cell gene expression data analysis},
  author={Wolf, F Alexander and Angerer, Philipp and Theis, Fabian J},
  journal={Genome Biology},
  volume={19},
  number={1},
  pages={15},
  year={2018},
  % doi = {10.1186/s13059-017-1382-0},
  publisher={Springer}
}

@article{mearns2020deconstructing,
  title     = {Deconstructing Hunting Behavior Reveals a Tightly Coupled Stimulus-Response Loop},
  author    = {Mearns, Duncan S. and Donovan, Joseph C. and Fernandes, Ant{\'o}nio M. and Semmelhack, Julia L. and Baier, Herwig},
  journal   = {Current Biology},
  volume    = {30},
  number    = {1},
  pages     = {54--69.e9},
  year      = {2020},
  publisher = {Elsevier},
  %doi       = {10.1016/j.cub.2019.11.022}
}

@article{schneider2023learnable,
  title     = {Learnable latent embeddings for joint behavioural and neural analysis},
  author    = {Schneider, Steffen and Lee, Jin Hwa and Mathis, Mackenzie Weygandt},
  journal   = {Nature},
  volume    = {617},
  number    = {7960},
  pages     = {360--368},
  year      = {2023},
  publisher = {Nature Publishing Group},
  %doi       = {10.1038/s41586-023-06031-6}
}

@article{findling2025prior,
  title     = {Brain-wide representations of prior information in mouse decision-making},
  author    = {Findling, Charles and Hubert, F{\'e}lix and {International Brain Laboratory} and Acerbi, Luigi and Benson, Brandon and Benson, Julius and Birman, Daniel and Bonacchi, Niccol{\`o} and Buchanan, E. Kelly and Bruijns, Sebastian and Carandini, Matteo and others},
  journal   = {Nature},
  volume    = {645},
  number    = {8079},
  pages     = {192--200},
  year      = {2025},
  publisher = {Nature Publishing Group}
  %doi       = {10.1038/s41586-025-09226-1}
}

@techreport{gpt-55,
  author       = {OpenAI},
  title        = {{GPT-5.5 System Card}},
  institution  = {OpenAI},
  year         = {2026},
  note         = {Accessed: 2026-06-29},
  url          = {https://openai.com/index/gpt-5-5-system-card/}
}

@techreport{claude-opus-46,
  author       = {Anthropic},
  title        = {{Claude Opus 4.6 System Card}},
  institution  = {Anthropic},
  year         = {2026},
  note         = {Accessed: 2026-06-29},
  url          = {https://www.anthropic.com/claude-opus-4-6-system-card}
}

@techreport{claude-opus-47,
  author       = {Anthropic},
  title        = {{Claude Opus 4.7 System Card}},
  institution  = {Anthropic},
  year         = {2026},
  note         = {Accessed: 2026-06-29},
  url          = {https://www.anthropic.com/claude-opus-4-7-system-card}
}

@techreport{gemini-31-pro,
  author       = {Google},
  title        = {{Gemini 3.1 Pro Model Card}},
  institution  = {Anthropic},
  year         = {2026},
  note         = {Accessed: 2026-06-29},
  url          = {https://storage.googleapis.com/deepmind-media/Model-Cards/Gemini-3-1-Pro-Model-Card.pdf}
}

@article{Tanaka2026Plastic,
  title   = {Plastic landmark anchoring in zebrafish compass neurons},
  author  = {Tanaka, Ryosuke and Portugues, Ruben},
  journal = {Nature},
  volume  = {650},
  number  = {8102},
  pages   = {673--680},
  year    = {2026},
  %doi     = {10.1038/s41586-025-09888-x}
}

@article{Atanas2023BrainWide,
  title   = {Brain-wide representations of behavior spanning multiple timescales and states in {C.~elegans}},
  author  = {Atanas, Adam A. and Kim, Jungsoo and Wang, Ziyu and others},
  journal = {Cell},
  volume  = {186},
  number  = {19},
  pages   = {4134--4151.e31},
  year    = {2023},
  %doi     = {10.1016/j.cell.2023.07.035}
}

@article{MarquezLegorreta2026WholeBrain,
  title   = {Whole-Brain Co-Mapping of Gene Expression and Neuronal Activity at Cellular Resolution in Behaving Zebrafish},
  author  = {Marquez-Legorreta, Emmanuel and Fleishman, Greg M. and Hesselink, Luuk W. and others},
  journal = {bioRxiv},
  year    = {2026},
  month   = feb,
  %doi     = {10.64898/2026.02.07.704095}
}

@article{Basnak2025Multimodal,
  title   = {Multimodal cue integration and learning in a neural representation of head direction},
  author  = {Basnak, Melanie A. and Kutschireiter, Anna and Okubo, Tatsuo S. and Chen, Albert and Gorelik, Pavel and Drugowitsch, Jan and Wilson, Rachel I.},
  journal = {Nature Neuroscience},
  volume  = {28},
  number  = {8},
  pages   = {1729--1740},
  year    = {2025},
  month   = jul,
  %doi     = {10.1038/s41593-024-01823-z},
  %url     = {https://doi.org/10.1038/s41593-024-01823-z}
}

@article{Gao2025Continuous,
  title   = {Continuous cell-type diversification in mouse visual cortex development},
  author  = {Gao, Yuan and van Velthoven, Cindy T. J. and Lee, Changkyu and others},
  journal = {Nature},
  volume  = {647},
  number  = {8088},
  pages   = {127--142},
  year    = {2025},
  %doi     = {10.1038/s41586-025-09644-1}
}

@article{Genkin2025Dynamics,
  title   = {The dynamics and geometry of choice in the premotor cortex},
  author  = {Genkin, Mikhail and Shenoy, Krishna V. and Chandrasekaran, Chandramouli and Engel, Tatiana A.},
  journal = {Nature},
  volume  = {645},
  number  = {8079},
  pages   = {168--176},
  year    = {2025},
  %doi     = {10.1038/s41586-025-09199-1}
}

@article{Yu2025Cholinergic,
  title   = {Cholinergic feedback for modality- and context-specific modulation of sensory representations},
  author  = {Yu, Bin and Yue, Yuxuan and Ren, Chi and Yun, Rui and Lim, Byungkook and Komiyama, Takaki},
  journal = {Science},
  volume  = {388},
  number  = {6753},
  pages   = {1324--1329},
  year    = {2025},
  %doi     = {10.1126/science.ads9152},
}

@article{Legare2025Structural,
  title   = {Structural and genetic determinants of zebrafish functional brain networks},
  author  = {L{\'e}gar{\'e}, Antoine and Lemieux, Mado and Boily, Vincent and Poulin, Sandrine and L{\'e}gar{\'e}, Arthur and Desrosiers, Patrick and De Koninck, Paul},
  journal = {Science Advances},
  volume  = {11},
  number  = {28},
  pages   = {eadv7576},
  year    = {2025},
  %doi     = {10.1126/sciadv.adv7576}
}

@article{ding2025scitoolagent,
  title={SciToolAgent: a knowledge-graph-driven scientific agent for multitool integration},
  author={Ding, Keyan and Yu, Jian and Huang, Jiaxin and Yang, Yuxin and Zhang, Qi and Chen, Huajun},
  journal   = {Nature Computational Science},
  volume    = {5},
  %number    = {10},
  pages     = {962--972},
  year      = {2025},
  publisher = {Nature Publishing Group}
  % doi       = {10.1038/s43588-025-00849-y}
}

@article{wang2025geneagent,
  title={GeneAgent: self-verification language agent for gene-set analysis using domain databases},
  author    = {Wang, Zhizheng and Jin, Qiao and Wei, Chih-Hsuan and Tian, Shubo and Lai, Po-Ting and Zhu, Qingqing and Day, Chi-Ping and Ross, Christina and Leaman, Robert and Lu, Zhiyong},
  journal={Nature Methods},
  volume={22},
  pages={1677--1685},
  year={2025},
  %doi={10.1038/s41592-025-02748-6},
  publisher={Nature Publishing Group}
}

@article{tang2025risks,
  title={Risks of AI scientists: prioritizing safeguarding over autonomy},
  author    = {Tang, Xiangru and Jin, Qiao and Zhu, Kunlun and Yuan, Tongxin and Zhang, Yichi and Zhou, Wangchunshu and Qu, Meng and Zhao, Yilun and Tang, Jian and Zhang, Zhuosheng and Cohan, Arman and Greenbaum, Dov and Lu, Zhiyong and Gerstein, Mark},
  journal   = {Nature Communications},
  volume    = {16},
  %number    = {1},
  pages     = {8317},
  year      = {2025},
  publisher = {Nature Publishing Group},
  %doi={10.1038/s41467-025-63913-1}, 
}

@article{messeri2024illusions,
  title={Artificial intelligence and illusions of understanding in scientific research},
  author    = {Messeri, Lisa and Crockett, M. J.},
  journal   = {Nature},
  volume    = {627},
  %number    = {8002},
  pages     = {49--58},
  year      = {2024},
  publisher = {Nature Publishing Group}
  %doi       = {10.1038/s41586-024-07146-0}
}

@article{hle,
  title     = {A benchmark of expert-level academic questions to assess AI capabilities},
  author    = {{Center for AI Safety} and {Scale AI} and {HLE Contributors Consortium}},
  journal   = {Nature},
  volume    = {649},
  number    = {8099},
  pages     = {1139--1146},
  year      = {2026},
  publisher = {Nature Publishing Group},
  %doi       = {10.1038/s41586-025-09962-4},
}

@article{swebench,
  title   = {{SWE-bench}: Can Language Models Resolve Real-World {GitHub} Issues?},
  author  = {Jimenez, Carlos E. and Yang, John and Wettig, Alexander and Yao, Shunyu and Pei, Kexin and Press, Ofir and Narasimhan, Karthik},
  journal = {arXiv preprint arXiv:2310.06770},
  year    = {2023},
  % doi     = {10.48550/arXiv.2310.06770},
  % url     = {https://arxiv.org/abs/2310.06770}
}

@article{mlebench,
  title   = {{MLE-bench}: Evaluating Machine Learning Agents on Machine Learning Engineering},
  author  = {Chan, Jun Shern and Chowdhury, Neil and Jaffe, Oliver and Aung, James and Sherburn, Dane and Mays, Evan and Starace, Giulio and Liu, Kevin and Maksin, Leon and Patwardhan, Tejal and Weng, Lilian and M{\k{a}}dry, Aleksander},
  journal = {arXiv preprint arXiv:2410.07095},
  year    = {2024},
  %doi     = {10.48550/arXiv.2410.07095}
}

@article{stringer2019spontaneous,
  title     = {Spontaneous behaviors drive multidimensional, brainwide activity},
  author    = {Stringer, Carsen and Pachitariu, Marius and Steinmetz, Nicholas and Reddy, Charu Bai and Carandini, Matteo and Harris, Kenneth D.},
  journal   = {Science},
  volume    = {364},
  number    = {6437},
  pages     = {eaav7893},
  year      = {2019},
  publisher = {American Association for the Advancement of Science},
  %doi       = {10.1126/science.aav7893}
}

@article{musall2019singletrial,
  title     = {Single-trial neural dynamics are dominated by richly varied movements},
  author    = {Musall, Simon and Kaufman, Matthew T. and Juavinett, Ashley L. and Gluf, Steven and Churchland, Anne K.},
  journal   = {Nature Neuroscience},
  volume    = {22},
  number    = {10},
  pages     = {1677--1686},
  year      = {2019},
  publisher = {Nature Publishing Group},
  %doi       = {10.1038/s41593-019-0502-4}
}

@article{steinmetz2019distributed,
  title     = {Distributed coding of choice, action and engagement across the mouse brain},
  author    = {Steinmetz, Nicholas A. and Zatka-Haas, Peter and Carandini, Matteo and Harris, Kenneth D.},
  journal   = {Nature},
  volume    = {576},
  number    = {7786},
  pages     = {266--273},
  year      = {2019},
  publisher = {Nature Publishing Group},
  %doi       = {10.1038/s41586-019-1787-x}
}

@article{rubel2022nwb,
  title   = {The {N}eurodata {W}ithout {B}orders ecosystem for neurophysiological data science},
  author  = {R{\"u}bel, Oliver and Tritt, Andrew and Ly, Ryan and others},
  journal = {eLife},
  volume  = {11},
  pages   = {e78362},
  year    = {2022},
  %doi     = {10.7554/eLife.78362}
}

@article{feng2026internagent,
  title={InternAgent-1.5: A Unified Agentic Framework for Long-Horizon Autonomous Scientific Discovery},
  author={Shiyang Feng and Runmin Ma and Xiangchao Yan and Yue Fan and Yusong Hu and Songtao Huang and others},
  journal={arXiv preprint arXiv:2602.08990},
  year={2026}
}

@misc{internagent-a1,
      title={Scaling the Horizon, Not the Parameters: Reaching Trillion-Parameter Performance with a 35B Agent}, 
      author={Lei Bai and Zongsheng Cao and Yang Chen and Zhiyao Cui and Shangheng Du and Yue Fan and Shiyang Feng and Zijie Guo and Haonan He and Liang He and Xiaohan He and Shuyue Hu and Yusong Hu and Songtao Huang and Yichen Jiang and Hao Li and Xin Li and Dahua Lin and Weihao Lin and Fenghua Ling and Dongrui Liu and Zhuo Liu and Runmin Ma and Chunjiang Mu and Haoyang Peng and Tianshuo Peng and Jinxin Shi and Luohe Shi and Boyuan Sun and Zelin Tan and Shengji Tang and Qianyi Wang and Yiming Wu and Yi Xie and Xiangchao Yan and Jingqi Ye and Peng Ye and Fangchen Yu and Jiakang Yuan and Bihao Zhan and Bo Zhang and Chen Zhang and Shufei Zhang and Shuaiyu Zhang and Wenlong Zhang and Yiqun Zhang and Junpeng Zhao and Zhijie Zhong and Bowen Zhou and Yuhao Zhou},
      year={2026},
      eprint={2606.30616},
      archivePrefix={arXiv},
      primaryClass={cs.CL},
      url={https://arxiv.org/abs/2606.30616}, 
}

@article{team2025internagent,
  title={InternAgent: When Agent Becomes the Scientist--Building Closed-Loop System from Hypothesis to Verification},
  author={Team, InternAgent and Zhang, Bo and Feng, Shiyang and Yan, Xiangchao and Yuan, Jiakang and Ma, Runmin and Hu, Yusong and Yu, Zhiyin and He, Xiaohan and Huang, Songtao and others},
  journal={arXiv preprint arXiv:2505.16938},
  year={2025}
}

@article{Luo2025,
  author    = {Luo, Xiaoliang and Rechardt, Akilles and Sun, Guangzhi and Nejad, Kevin K. and Y{\'a}{\~n}ez, Felipe and Yilmaz, Bati and Lee, Kangjoo and Cohen, Alexandra O. and Borghesani, Valentina and Pashkov, Anton and Marinazzo, Daniele and Nicholas, Jonathan and Salatiello, Alessandro and Sucholutsky, Ilia and Minervini, Pasquale and Razavi, Sepehr and Rocca, Roberta and Yusifov, Elkhan and Okalova, Tereza and Gu, Nianlong and Ferianc, Martin and Khona, Mikail and Patil, Kaustubh R. and Lee, Pui-Shee and Mata, Rui and Myers, Nicholas E. and Bizley, Jennifer K. and Musslick, Sebastian and Bilgin, Isil Poyraz and Niso, Guiomar and Ales, Justin M. and Gaebler, Michael and Ratan Murty, N. Apurva and Loued-Khenissi, Leyla and Behler, Anna and Hall, Chloe M. and Dafflon, Jessica and Bao, Sherry Dongqi and Love, Bradley C.},
  title     = {Large language models surpass human experts in predicting neuroscience results},
  journal   = {Nature Human Behaviour},
  year      = {2025},
  volume    = {9},
  number    = {2},
  pages     = {305--315},
  %doi       = {10.1038/s41562-024-02046-9},
  issn      = {2397-3374}
}

@article{Marwitz2026,
  author    = {Marwitz, Thomas and Colsmann, Alexander and Breitung, Ben and Brabec, Christoph and Kirchlechner, Christoph and Blasco, Eva and Marques, Gabriel Cadilha and Hahn, Horst and Hirtz, Michael and Levkin, Pavel A. and Eggeler, Yolita M. and Schl{\"o}der, Tobias and Friederich, Pascal},
  title     = {Predicting new research directions in materials science using large language models and concept graphs},
  journal   = {Nature Machine Intelligence},
  year      = {2026},
  volume    = {8},
  number    = {4},
  pages     = {535--544},
  %doi       = {10.1038/s42256-026-01206-y},
  %issn      = {2522-5839}
}
\endgroup

\clearpage
\appendix

\setcounter{table}{0}
\setcounter{figure}{0}
\renewcommand{\thetable}{\arabic{table}}
\renewcommand{\thefigure}{\arabic{figure}}
\captionsetup[table]{name=Supplementary Table}
\captionsetup[figure]{name=Supplementary Figure}

\section{Supplementary Information}
\label{sec:appendix}

\subsection{Supplementary Notes}

\subsubsection{Implementation details of SeekBrain}

This Supplementary Note provides the implementation details of analysis execution engine. For each task, SeekBrain receives a natural-language neuroscience question together with optional dataset metadata and data-path information. The system then (1) retrieves a task-relevant analysis recipe from the Neuroscience Analysis Repertoire, (2) formulates a data-grounded analysis schema, (3) generates and executes Python code, (4) validates the output, and (5) interprets the validated figures, statistics, and execution traces to produce a self-contained scientific conclusion.

In the configuration used for all reported experiments, analytical formulation, code execution, validation, and result interpretation are routed to Claude Opus 4.7. The architecture supports modular replacement of base LLMs for each agent.

\paragraph{Analysis execution workflow.}

Before analytical formulation, the workflow retrieves a targeted recipe from the Neuroscience Analysis Repertoire. If no verified recipe is matched, the workflow initiates a task-specific recipe crystallization process: it identifies relevant publications with associated code repositories, selects task-relevant paper and repository context, and crystallizes the paper-code correspondence into a reusable analysis recipe. The retrieved or newly generated recipe is then supplied to the Formulation Agent as a methodological reference.

The Formulation Agent produces a data-grounded methodological schema by integrating the user question, dataset metadata, recipe, and literature background. The Code Agent writes, executes, and debugs Python analysis code within a bounded working directory, producing 
required code, outputting the generated scripts, execution logs, figures, statistics, and execution logs. The workflow explicitly prohibits the use of dummy, placeholder, or proxy data. Instead, missing data errors are either resolved via user-supplied paths or formally recorded as limitations. 
The Validation Agent evaluates whether the output answers the original question and conforms to the schema, drawing on the figure, statistics, optional reference figure, code-agent confidence tag, and literature-grounded background. 
The Validation Agent can reject a result by flagging specific mismatches. This feedback is routed back to the Code Agent for a maximum of three validation cycles before final interpretation.
Finally, the Interpretation Agent reports scientific conclusions extracted from the result figure, quantitative statistics, execution log, methodological schema, plausibility warnings, and literature background.

\paragraph{Analysis recipe grounding.}
Within the analysis workflow, the Neuroscience Analysis Repertoire functions as procedural memory. Any retrieved recipe serves as a methodological reference rather than a strict directive. The Formulation Agent evaluates its alignment with the current question, data modality, and validation requirements before deciding whether to adopt, adapt, or discard it. This approach ensures that the user's question and dataset metadata remain the primary constraints, while literature-validated procedures provide adaptable guidance.

\subsection{Supplementary Figures}
\label{app:prompts}

\begin{figure}[H]
    \centering
    \includegraphics[width=\textwidth]{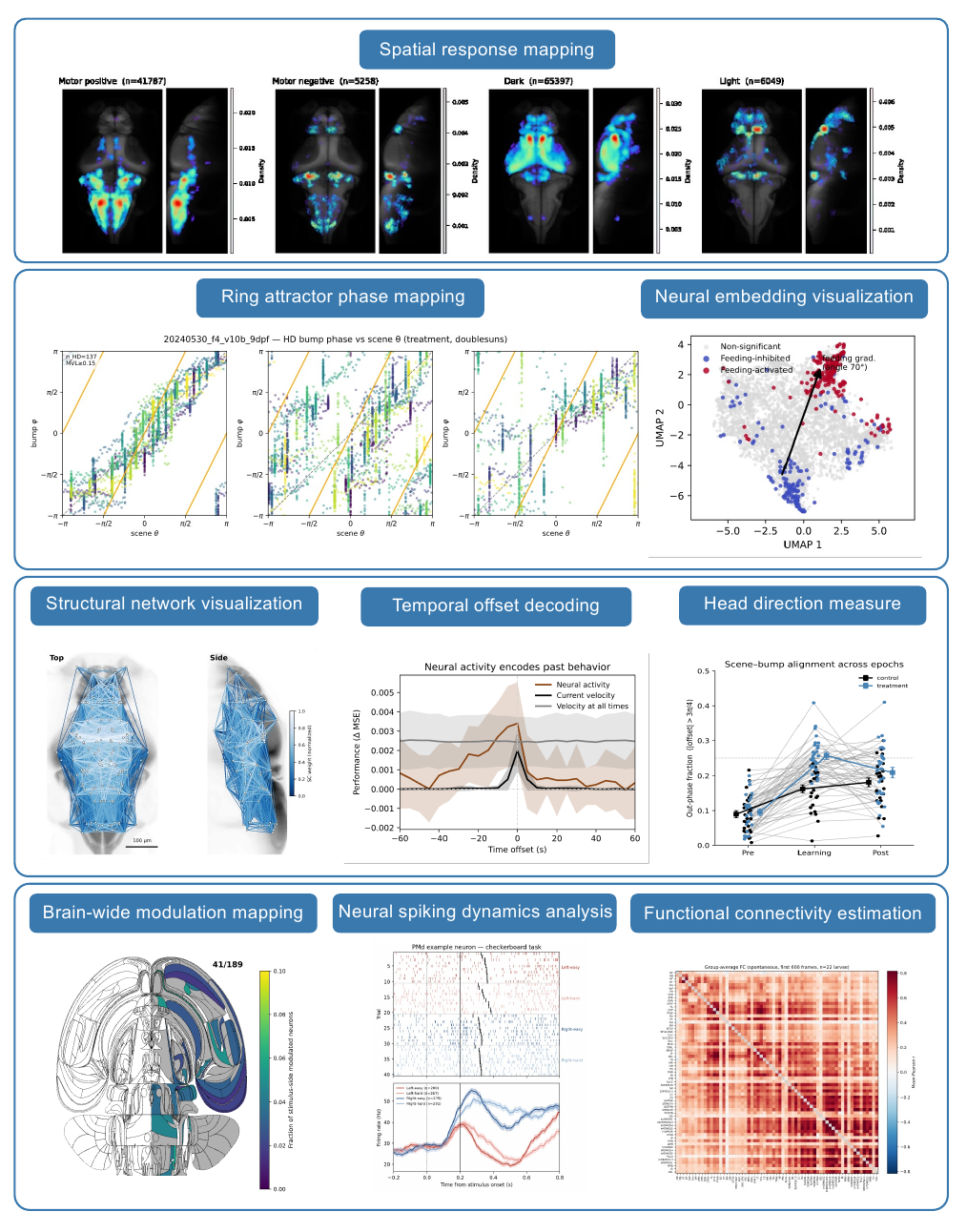}
    \caption{Examples of SeekBrain's analysis results on BrainArena.}
    \label{benchmark-cases}
\end{figure}

\begin{figure}[htbp]
    \centering
    \includegraphics[width=1.0\textwidth]{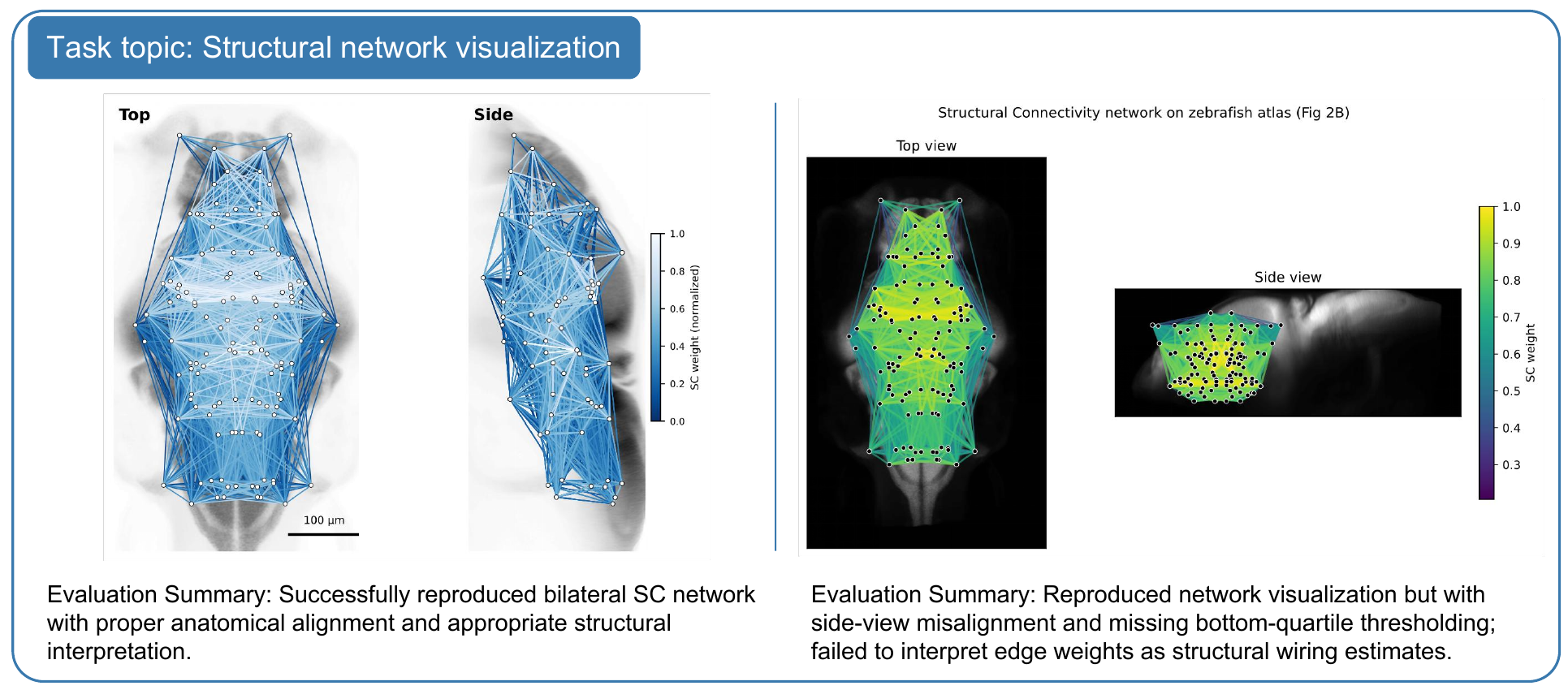}
    \caption{Cases with and without established analysis recipes.}
    \label{skill-cases}
    \vspace{6mm}
\end{figure}

\NewTColorBox{EqBox}{ O{} m }{%
  breakable,
  enhanced,
  colback=white,
  colframe=black!70,
  coltitle=white,
  colbacktitle=black!70,
  title={\textsc{#2}},
  fonttitle=\bfseries,
  left=1.2mm,
  right=1.2mm,
  top=1.2mm,
  bottom=1.2mm,
  boxrule=0.6pt,
}
\newcommand{\PromptRole}[1]{\par\medskip\noindent{\ttfamily\small\bfseries #1}\par\medskip}

\newcommand{\PromptDivider}{\par\smallskip\noindent{\color{gray}\rule{\linewidth}{0.2pt}}\par\smallskip}

\newcommand{\PromptText}{\rmfamily\footnotesize\obeylines}

\begin{EqBox}[!htbp]{Prompt for Benchmark Task Execution}
\vspace{1mm}
{\PromptText
Your task is to complete a scientific data analysis task based on the provided specifications.

\vspace{0.2cm}
\textbf{Task Parameters:}
\begin{itemize}
    \setlength{\itemsep}{0pt}
    \setlength{\parskip}{0pt}
    \item Question file: \texttt{\{QUESTION\_FILE\}}
    \item Data directory: \texttt{\{DATA\_DIR\}}
\end{itemize}

\vspace{0.2cm}
\textbf{Output Requirements:}
Save all required outputs to the \texttt{\{OUTPUT\_DIR\}} directory:
\begin{enumerate}
    \setlength{\itemsep}{0pt}
    \setlength{\parskip}{0pt}
    \item \textbf{\texttt{conclusion.md}}: Write 1 to 3 scientific conclusions. Each conclusion must be a single, self-contained sentence or short paragraph focusing on key scientific findings, supported by quantitative evidence where available.
    \item \textbf{\texttt{analyze.md}}: Document the complete analysis process, including data exploration, methodology, result interpretation, and your reasoning trace.
    \item \textbf{\texttt{analyze.png}}: Generate a figure that closely matches the visual and analytical specifications in the question file.
    \item \textbf{\texttt{analyze.py}}: Provide the complete executable analysis code script.
\end{enumerate}

\vspace{0.2cm}
\textbf{Execution Strategy:}
\begin{itemize}
    \setlength{\itemsep}{0pt}
    \setlength{\parskip}{0pt}
    \item First, explore the data structure. Next, write and execute the analysis code. Finally, draft \texttt{analyze.md} and \texttt{conclusion.md}.
    \item Prioritize executing an end-to-end pipeline over exhaustive data exploration.
    \item Ensure the figure, analysis document, and conclusions are generated before the timeout, even if the results are partial or imperfect.
    \item Begin execution immediately without asking for user confirmation.
\end{itemize}
}
\end{EqBox}
\captionof{figure}{\textbf{Prompt for benchmark task execution.} Each coding agent receives this prompt alongside a task-specific question file and data directory, instructing it to perform data exploration, write analysis code, generate a figure, and synthesize scientific conclusions within a strict timeout}
\vspace{2mm}
\label{fig:prompt_task_execution}

\begin{EqBox}[!htbp]{Prompt for LLM-as-Judge Evaluation}
\vspace{1mm}
{\PromptText
You are a strict but fair evaluator for a scientific benchmark. Your task is to evaluate the submitted neuroscience benchmark results against the provided rubric. Grade only the submitted artifacts. Do not credit claims or figure content that are absent from the submission. Do not penalize phrasing differences if the scientific meaning is correct.

\vspace{0.2cm}
\textbf{Inputs Provided:}
\begin{enumerate}
    \setlength{\itemsep}{0pt}
    \setlength{\parskip}{0pt}
    \item \textbf{Ground-truth figure} from the original paper (Image)
    \item \textbf{Model's reproduced figure} (Image)
    \item \textbf{Rubric} (JSON): Ground-truth conclusion, details scoring criteria, point allocations (\texttt{max\_point}), and deduction rules (\texttt{deduction\_rule})
    \item \textbf{Model's reproduced code} (Python): The executable analysis and visualization code used to generate the reproduced figure
    \item \textbf{Model's submission} (Text): The conclusion or methods summary generated by the model
\end{enumerate}

\vspace{0.2cm}
\textbf{Grading Rules:}
\begin{itemize}
    \setlength{\itemsep}{0pt}
    \setlength{\parskip}{0pt}
    \item The maximum total score is \texttt{\{max\_score\}}. Assign each rubric item a score from 0 up to its specified \texttt{max\_point}.
    \item Award partial credit if the submission partially satisfies a criterion.
    \item Apply each \texttt{deduction\_rule} strictly when applicable.
    \item Treat the rubric's ground-truth conclusion as the scoring reference, not as text automatically credited to the submission.
    \item Scores may be integers or decimals.
\end{itemize}

\vspace{0.2cm}
\textbf{Task \& Output Format:}
Score the model's conclusion from 0 to 100 using the rubric. Provide your step-by-step reasoning and the final score. Return ONLY valid JSON with no Markdown fences or additional text, adhering strictly to this schema:

\vspace{0.1cm}
\noindent\texttt{\{\\}
\texttt{~~"reasoning": "<your detailed step-by-step reasoning>",\\}
\texttt{~~"score": <final numerical score>\\}
\texttt{\}}
}
\end{EqBox}
\captionof{figure}{\textbf{Prompt for LLM-as-Judge evaluation.} The evaluator receives multimodal inputs: reproduced codes, reproduced figures, structured analysis metadata, a detailed rubric, and the model's written conclusions.} It then assigns a score from 0 to 100 following strict grading rules with partial credit and specified deduction policies.
\label{fig:prompt_evaluation}

\begin{EqBox}[!htbp]{Example of a Neuroscience Analysis Recipe}
\vspace{1mm}
{\PromptText
\textbf{Recipe File Tree:}

\vspace{0.1cm}
{\ttfamily\small
.\par
|-- RECIPE.md\par
|-- metadata.json\par
|-- references\par
|\phantom{--}|-- implementation\_notes.md\par
|\phantom{--}`-- scientific\_context.md\par
`-- scripts\par
\phantom{----}|-- run\_analysis.py\par
\phantom{----}`-- utils.py
}

\vspace{0.2cm}
\textbf{Key Contents of \texttt{RECIPE.md}:}
\begin{itemize}
    \setlength{\itemsep}{2pt}
    \setlength{\parskip}{0pt}
    \item \textbf{Objective:} Analyze how head-direction cell tuning transforms when a zebrafish experiences a symmetric virtual scene. Align tuning curves to each neuron's pre-learning preferred orientation and compare pre-learning, learning, and post-learning epochs to identify systematic bimodal or double-peaked responses caused by visual ambiguity.
    \item \textbf{Input:} Synchronized behavioral data containing scene orientation $\theta$ and epoch boundaries; neural data containing ROI-matched $\Delta F/F$ traces; stable \texttt{ROI\_IDs} across all three epochs; and MATLAB circular-statistics and plotting utilities, including \texttt{Kent\_func}, \texttt{stats\_plot}, and \texttt{im\_dissim}.
    \item \textbf{Processing:} Identify head-direction cells in the pre-learning epoch using a Rayleigh significance criterion ($p<0.05$), estimate each cell's preferred orientation $b$, and center its tuning curve with $\theta-b$ wrapped to $[-\pi,\pi]$. Sort ROIs once by pre-learning $b$ and preserve this order across all epochs. Construct three tuning-curve heatmaps with a shared color scale, then fit a two-component mixture model to learning-epoch curves to quantify the secondary peak near $\pm\pi$.
    \item \textbf{Output:} (1) A synchronized three-panel heatmap for the Pre, Learning, and Post epochs; (2) \texttt{centered\_tuning\_matrices.mat}; (3) \texttt{preferred\_orientations.mat}; (4) \texttt{ROI\_order.mat}; and (5) mixture-model statistics describing bimodality. A valid result preserves identical ROI ordering and color limits across panels and reveals whether symmetric-scene learning produces a population-level second tuning peak.
\end{itemize}
}
\end{EqBox}
\captionof{figure}{\textbf{Example of a neuroscience analysis recipe.} The complete skill directory tree is shown first, followed by the essential objective, input specification, and output specification.}
\label{fig:recipe_example_zebrafish_hd_tuning}

\subsection{Supplementary Tables}

\begingroup
\footnotesize
\setlength{\tabcolsep}{2.5pt}
\renewcommand{\arraystretch}{1.12}
\setlength{\emergencystretch}{2em}

\setlength{\LTleft}{0pt}
\setlength{\LTright}{\fill}

\begin{longtable}{
  @{}
  >{\RaggedRight\arraybackslash}p{0.29\textwidth}
  >{\RaggedRight\arraybackslash}p{0.50\textwidth}
  >{\RaggedRight\arraybackslash}p{0.17\textwidth}
  @{}
}
\caption{Neuroscience analysis recipes.}
\label{tab:neuroscience_analysis_recipes}\\
\toprule
\textbf{Name} & \textbf{Description} & \textbf{Modality} \\
\midrule
\endfirsthead

\toprule
\textbf{Name} & \textbf{Description} & \textbf{Modality} \\
\midrule
\endhead

ana-mouse-behavior-markov-epoch\_interval & Decomposes trial-by-trial choices into interpretable latent behavioral strategies using a logistic-emission GLM-HMM fitted by expectation-maximization and decoded with Viterbi inference. It produces state-specific GLM weights, posterior state-probability traces, decoded sequences, and AIC/BIC model-selection scores. & Behavioral Modality  \\
ana-mouse-behavior-glm-performance & Fits Bayes-optimal, stimulus-kernel, and action-kernel models to trial-by-trial choice data, estimates session-level log evidence, and applies random-effects Bayesian model selection. It outputs model frequencies, exceedance probabilities, evidence matrices, and summary comparisons of competing behavioral strategies. & Behavioral Modality  \\
ana-mouse-behavior-anova-task\_performance & Analyzes trial-aligned behavioral and calcium-imaging data using OASIS deconvolution, learning-curve fitting, and groupwise statistical tests. It produces performance time courses and significance estimates for CNO-induced olfactory learning impairment and post-washout recovery. & Behavioral Modality  \\
multi-mouse-ca-odor\_tuning\_ratio\_chi\_square-pie\_plot & Pools binary response masks to calculate percentages of excited, suppressed, and non-responsive granule cell–odor pairs under task and passive conditions. It outputs summary statistics and comparative side-by-side pie charts to reveal sparse, state-dependent odor-response distributions. & Dynamics Modality  \\
ana-zebrafish-ca-circular\_null-cofluctuation\_event\_trace & Detects statistically significant brain-wide co-fluctuation events in zebrafish calcium data using edge-time-series RMS, region-wise circular-shift null distributions, and Benjamini–Hochberg FDR. It outputs event-level regional activity vectors for downstream brain-state clustering, plus RMS and null statistics. & Dynamics Modality  \\
ana-zebrafish-ca-cross\_day\_correlation-identity\_fingerprint\_matrix & Tests cross-day zebrafish larva re-identification by correlating vectorized upper triangles of functional-connectivity matrices and symmetrizing pairwise similarities. It outputs identification accuracy, a larva-by-larva similarity matrix, and a heatmap marking each larva’s strongest match. & Dynamics Modality  \\
ana-zebrafish-ca-fingerprint\_mannwhitney-box\_plot & Tests zebrafish functional-connectivity fingerprint significance by comparing cross-day intra- and inter-individual similarities with a two-sided Mann–Whitney U test. It outputs the similarity matrix and group vectors plus a jittered min-to-max boxplot annotated with the p-value. & Dynamics Modality  \\
ana-zebrafish-ca-pearson-fc\_matrix & Builds per-larva and group-average zebrafish functional connectivity matrices by Gaussian-smoothing regional $\Delta F / F_{0}$ traces and computing pairwise Pearson correlations after quality-control exclusions. It outputs region-by-region matrices quantifying brain-wide functional coupling and synchronization. & Dynamics Modality  \\
ana-zebrafish-ca\_fc\_matrix-cross\_individual\_correlation-network\_stereotypy\_matrix & Quantifies inter-larva functional-connectivity stereotypy by correlating vectorized upper triangles of individual zebrafish FC matrices. It outputs an ordered larva-by-larva Pearson similarity matrix, representative-larva ranking, summary statistics, and lower-triangular heatmap. & Dynamics Modality  \\
pre-zebrafish-ca-detrend\_minfilter\_zscore-activity\_state\_matrix & Builds a cross-fish bilateral zebrafish brain-state matrix by symmetrically excluding unusable regions, applying min-filter $\Delta F / F_{0}$ detrending, and z-scoring regional calcium traces before time-wise concatenation. It produces standardized cohort activity arrays for heatmaps, co-fluctuation detection, and brain-state clustering. & Dynamics Modality  \\
pre-zebrafish-ca-baseline\_correlation\_minfilter-dff\_region\_timeseries & Converts motion-corrected zebrafish fluorescence to per-neuron $\Delta F / F_{0}$ using a Gaussian-smoothed running-minimum baseline, then averages activity across atlas regions. It saves neuronal and regional time series and produces an anterior-to-posterior stacked regional-trace plot with the stimulus window marked. & Dynamics Modality  \\
ana-zebrafish-ca-tsne\_hdbscan-correlation\_matrix\_heatmap & Clusters high-amplitude zebrafish whole-brain coactivation events using t-SNE and HDBSCAN, reassigning noise by correlation to native-space cluster centroids. It outputs discrete brain-state labels, embeddings, and a cluster-sorted event correlation heatmap revealing block-diagonal state structure. & Dynamics Modality  \\
vis-zebrafish-fc\_atlas-edge\_projection-network\_map & Projects a group-averaged Pearson functional connectivity matrix onto Mapzebrain top and side anatomy using bilateral regional centroids as nodes and correlation-weighted edges, hiding the weakest quartile. It produces a publication-quality spatial network figure for assessing the anatomical distribution of strong connections. & Dynamics Modality  \\
ana-zebrafish-ca-eigen\_spectrum-fc\_scree\_plot & Visualizes the diffusion-map eigenvalue spectrum of zebrafish whole-brain functional connectivity as a scree plot. It reveals spectral decay and the elbow supporting retention of leading functional gradients, including the 50-mode cutoff used for structural–functional gradient matching. & Dynamics Modality  \\
ana-zebrafish-ca-tsne\_hdbscan-cluster\_scatter\_plot & Embeds high-amplitude zebrafish whole-brain coactivation events with t-SNE, clusters them using HDBSCAN, and reassigns noise by correlation to centroids in the original regional space. It outputs cluster labels, embedding coordinates, and a brain-state-coloured 2-D scatter plot. & Dynamics Modality  \\
ana-mouse-omics-hierarchical\_clustering-type\_temporal\_trajectory & Quantifies developmental changes in scRNA-seq cell-type diversity using cross-age integration, resolution-consistent clustering, downsampling, and cluster stability analysis. It produces temporal plots and tables of cluster counts to identify emerging populations and diversification plateaus. & Molecular Modality  \\
ana-mouse-omics-differential\_gene-temporal\_expression\_profile & Quantifies subclass-specific developmental gene expression changes in mouse visual cortex using replicate-aware pseudobulk aggregation and DESeq2 or MAST. It produces per-gene log2 fold changes and adjusted significance values, DEG summaries, volcano plots, and up/downregulated gene counts across subclasses. & Molecular Modality  \\
ana-mouse-omics-gam-expression\_profile & Models lineage-specific marker expression across developmental time using generalized additive models and spline derivatives, classifying trajectories as Up, Down, Transient, or Constant. It outputs a classification table and faceted expression curves with confidence intervals for excitatory, inhibitory, and glial lineages. & Molecular Modality  \\
ana-mouse-omics-hierarchical\_clustering-scatter\_type\_temporal\_trajectory & Analyzes developmental scRNA-seq using integrated hierarchical clustering, temporal linkage, and Jaccard similarity to track cell-type emergence, branching, and stability. It outputs age-resolved subcluster counts, stability metrics, and alluvial diagrams of lineage trajectories. & Molecular Modality  \\
vis-zebrafish-gene-zscore-expression\_matrix\_heatmap & Visualizes larval zebrafish spatial transcriptomics as a grayscale heatmap of genes across retained brain regions, z-scoring genes within each region. It produces a figure and descriptive statistics to inspect regional marker patterns before gene–connectivity comparisons. & Molecular Modality  \\
ana-zebrafish-swc-nblast-morphological\_metric & Classifies zebrafish cerebellar efferent neurons by computing all-by-all NBLAST morphological similarities and applying Ward hierarchical clustering. It produces an eight-cluster dendrogram, neuron classification table, and silhouette scores distinguishing Purkinje and eurydendroid morphologies. & Structural Modality  \\
ana-zebrafish-connectome-pearson-corrcoef\_distance\_scatter\_plot & Quantifies distance-dependent connectivity in the larval zebrafish brain using Euclidean centroid distances, volume-normalized neurite lengths, and Pearson correlation. It also applies hierarchical consensus clustering to produce statistical evidence, a distance–strength plot, and a dendrogram of stable brain communities. & Structural Modality  \\
ana-zebrafish-swc-nblast-morphology & Computes all-by-all NBLAST similarities for registered SWC neuronal skeletons, applies hierarchical clustering to define morphological groups, and renders them in a standard 3D brain template. Outputs a dendrogram, neuron-to-cluster map, and cluster visualization gallery. & Structural Modality  \\
ana-zebrafish-sc\_matrix-hierarchical\_consensus\_clustering-consensus\_matrix\_dendrogram & Identifies zebrafish structural-connectivity modules by applying Ward hierarchical clustering to co-assignment probabilities from multi-resolution Louvain partitions. It outputs module labels reordered by anterior-posterior position, the co-assignment matrix, and a dendrogram-matrix visualization. & Structural Modality  \\
vis-zebrafish-sc\_atlas-edge\_projection-network\_map & Visualizes a bilateral zebrafish structural connectivity network by overlaying atlas-centroid nodes and thresholded, weight-coded edges on dorsal and sagittal MapzebrainAtlas projections. It outputs a dual-panel figure and JSON statistics summarizing regions, edges, weights, and coordinate ranges. & Structural Modality  \\
ana-zebrafish-structure\_connectivity-graph\_predictor\_pca-scree\_plot & Applies unscaled PCA to structural predictors derived from directed and undirected zebrafish connectomes and inter-regional Euclidean distances. It reports and plots per-component and cumulative explained variance for the first 10 PCs to assess predictor redundancy and dimensionality. & Structural Modality  \\
pre-zebrafish-cell\_coordinate\_anatomy-region\_census-atlas\_contour\_map\_cell\_density & Censuses neurons per larva across Mapzebrain regions and excludes retina plus regions with fewer than 25 neurons in at least half of larvae. It reports whole-brain counts, writes the downstream QC mask, and generates anatomical projection backgrounds. & Structural Modality  \\
ana-zebrafish-sc\_matrix\_anatomy-hierarchical\_consensus\_clustering-module\_topography & Projects consensus-Louvain structural-connectivity module labels onto bilateral zebrafish atlas centroids in color-coded top and side views. It produces an anatomical topography figure and module counts to show anterior–posterior organization and spatial contiguity. & Structural Modality  \\
ana-zebrafish-sc\_matrix-eigen\_spectrum-sc\_scree\_plot & Plots the larval zebrafish structural-connectivity diffusion-map eigenvalue spectrum as a scree curve to reveal its decay and elbow. It outputs a figure and summary statistics for the eigenvalue count, first value, mean, median, minimum, and maximum. & Structural Modality  \\
vis-zebrafish-sc-module\_sorted-sc\_block\_heatmap & Visualizes the zebrafish structural connectome by sorting its adjacency matrix by four consensus-Louvain modules, squaring weights for contrast, and outlining diagonal blocks. It produces a heatmap and descriptive statistics showing within- versus between-module connectivity structure. & Structural Modality  \\
multi-zebrafish-ca\_behavior-vonmises-phase\_locking & Quantifies zebrafish neural bump–scene phase coupling during jump and noise epochs by fitting von Mises concentration to angular offsets and comparing observed $\kappa$ values with circular-shift null distributions. It outputs per-fish coupling statistics, 95th-percentile significance thresholds, and summary plots. & Dynamics-Behavior Coupling  \\
multi-zebrafish-ca\_behavior-circular\_statistics-phase\_locking & Quantifies zebrafish scene–neural bump alignment disruption using circular angular differences and thresholding to calculate each fish’s out-of-phase fraction across pre-learning, learning, and post-learning epochs. It outputs per-subject fractions and a paired longitudinal trajectory plot. & Dynamics-Behavior Coupling  \\
multi-drosophila-ca\_behavior-circular\_statistics-encoding\_accuracy & Quantifies Drosophila E-PG head-direction encoding by preprocessing fluorescence data, estimating activity-bump phase, and calculating vector strength relative to heading. It statistically compares visual cue conditions and produces an SVG grouped scatter plot of per-fly accuracy, means, error bars, and significance. & Dynamics-Behavior Coupling  \\
multi-zebrafish-ca\_behavior-vonmises\_shuffling-distribution\_bar\_plot & Quantifies neural bump stability relative to visual scenes by fitting von Mises distributions to circular phase offsets and comparing observed concentration parameters with shuffled null distributions. It outputs epoch-level stability statistics and visualizations identifying significant phase-locking during perturbations. & Dynamics-Behavior Coupling  \\
multi-zebrafish-ca\_behavior-circular\_correlation-tuning\_curve & Analyzes circular correlations between decoded neural bump phase and scene orientation across pre-learning, learning, and post-training epochs. It outputs epoch-specific slopes and scatter plots to reveal shifts between 1:1 and 2:1 neural mappings. & Dynamics-Behavior Coupling  \\
multi-c\_elegans-ca\_behavior-umap-embedding\_scatter\_plot & Projects pooled multi-animal neural activity into a 2D UMAP and overlays CePNEM-derived feeding-tuning indices. It produces a neural-manifold plot and per-neuron coordinate table revealing graded transitions from feeding-inhibited to feeding-activated encoding. & Dynamics-Behavior Coupling  \\
multi-mouse-spike\_behavior-glm-decoding\_metrics & Decodes trial-by-trial prior probability from binned ORBvl inter-trial population activity using cross-validated linear models benchmarked against a Bayesian observer. It outputs decoded and reference prior traces plus session-level Pearson correlation and noise-ceiling-corrected R². & Dynamics-Behavior Coupling  \\
multi-c\_elegans-ca\_behavior-correlation-timeseries\_trace & Analyzes neuronal velocity encoding by correlating binned firing rates with causally EWMA-filtered velocity across decay constants. It outputs per-neuron correlation profiles, optimal integration timescales, significance against shuffled spike-time nulls, and representative plots. & Dynamics-Behavior Coupling  \\
multi-c\_elegans-ca\_behavior-half\_decay-distribution & Fits Ridge regression encoding kernels and exponential decays to estimate neuronal half-decay times, retaining neurons significant against a shuffle-based null. It outputs an ECDF of population timescales with ±1 SD inter-animal shading and per-neuron summary data. & Dynamics-Behavior Coupling  \\
multi-c\_elegans-ca\_behavior-ridge-timeseries\_trace & Uses time-lagged ridge regression to decode past, current, and future behavioral velocity from neural activity, comparing smoothed prediction errors against shuffled-label controls. It outputs performance across temporal offsets and plots mean decoding curves with variability across animals. & Dynamics-Behavior Coupling  \\
multi-zebrafish-ca\_behavior-circular\_statistics-tuning\_curve & Analyzes head-direction cell tuning across pre-learning, learning, and post-learning epochs using circular fitting, preferred-orientation alignment, and Gaussian-mixture bimodality assessment. It outputs synchronized, consistently sorted heatmaps and tuning-shift statistics revealing symmetry-induced double-peaked responses. & Dynamics-Behavior Coupling  \\
multi-macaque-spike\_behavior-glm-performance & Quantifies prior-dependent shifts in superior colliculus firing by epoching synchronized post-stimulus spikes and fitting logistic regressions with contrast, prior, and neural activity. It produces regression coefficients, prior-stratified neurometric curves and PSE shifts, and stimulus-aligned PSTHs. & Dynamics-Behavior Coupling  \\
multi-mouse-spike\_behavior-decay\_correlation-scatter\_plot & Fits exponential decays to neural latent-state autocorrelations and behavioral dwell times, then correlates their inverse time constants across quality-controlled, synchronized sessions. It produces a session-level scatter plot and a table of the correlation coefficient, p-value, and sample size. & Dynamics-Behavior Coupling  \\
multi-c\_elegans-ca\_behavior-pca-pc\_feature\_heatmap & Performs PCA on z-scored neural activity and Ward hierarchical clustering of neurons using the top five PC loadings. It aligns clusters with CePNEM behavioral tuning weights and produces a dendrogram and heatmaps revealing behavior-associated functional ensembles. & Dynamics-Behavior Coupling  \\
multi-macaque-spike\_behavior-classified\_firing\_rate-psth\_trace & Generates stimulus-aligned PSTHs for 14 PMd neurons using a 75 ms sliding window and 10 ms step, grouped by choice and difficulty. It outputs a multi-panel figure with trial-averaged firing rates and SEM shading to reveal condition-dependent temporal heterogeneity. & Dynamics-Behavior Coupling  \\
multi-drosophila-ca\_behavior-pearson-scatter\_plot & Analyzes movement-gated Drosophila EPG imaging and behavior data using circular phase decoding, von Mises bump-width fitting, and Pearson correlation. It outputs per-fly accuracy and width metrics, correlation statistics, and a regression scatter plot of heading-encoding accuracy versus bump width. & Dynamics-Behavior Coupling  \\
multi-macaque-spike\_behavior-classified\_firing\_rate-psth\_distribution & Computes stimulus-aligned, trial-averaged single-neuron PMd PSTHs by choice and difficulty using a 75-ms sliding boxcar, 10-ms steps, reaction-time masking, and SEM. It produces condition-wise plots and temporal-profile classifications to characterize ramping, transient, non-monotonic, and mixed decision-related activity. & Dynamics-Behavior Coupling  \\
multi-macaque-spike\_behavior-tensor\_decomposition-temporal\_motif & Applies CP tensor decomposition to PMd spike-rate data organized by neuron, time, and trial, then relates latent trial factors to choice and stimulus coherence. It outputs shared ramping or transient temporal motifs, behavioral sensitivities, and reconstructed population activity. & Dynamics-Behavior Coupling  \\
multi-macaque-spike\_behavior-classified\_firing\_rate-raster\_psth & Applies canonical polyadic decomposition to trial-binned PMd firing-rate tensors, extracting temporal factors and trial loadings linked to stimulus coherence and choice. It produces latent PSTH visualizations and condition-sorted loading plots validated against trial-averaged neural activity. & Dynamics-Behavior Coupling  \\
multi-mouse-spike\_behavior-classified\_firing\_rate-psth\_heatmap & Computes event-aligned population PSTHs by binning and Gaussian-smoothing spike trains, then averaging and optionally baseline-normalizing activity by stimulus difficulty and behavioral choice. It outputs condition-segregated firing-rate arrays and plots with SEM to reveal decision-related temporal dynamics. & Dynamics-Behavior Coupling  \\
multi-mouse-spike\_behavior-exponential\_kernel\_glm-decay\_distribution & Estimates session-specific trial-history decay constants from behavioral logistic kernels and neural GLMs, then measures their cross-session alignment using modality-stratified Pearson correlation. Outputs paired decay estimates, correlation statistics, and neural-versus-behavioral scatter plots. & Dynamics-Behavior Coupling  \\
multi-zebrafish-ca\_behavior-permutation-spatial\_density & Aggregates functionally classified neuron centroids across zebrafish in a common 3D atlas and uses permutation-based spatial p-values to identify locations consistent across individuals. It outputs filtered centroid and p-value arrays plus density-colored whole-brain maps for motor and visual response populations. & Dynamics-Behavior Coupling  \\
multi-zebrafish-ca\_behavior-linear\_regression-sensorimotor\_trace & Creates a single-larva sensorimotor figure aligning binarized swimming, light/dark stimuli, and calcium traces for the 50 most extreme neurons in four precomputed correlation-ranked ensembles. It overlays individual traces and ensemble means to qualitatively show motor- and stimulus-linked activity without population-level inference. & Dynamics-Behavior Coupling  \\
multi-zebrafish-ca\_behavior\_anatomy\_gene-correlation\_visual\_response-neuronal\_subtype\_features & Integrates atlas registration and NBLAST clustering with stimulus-aligned calcium analysis and neural–swimming correlations to characterize glyt2–gfra1a zebrafish neurons. It produces 3D brain maps, peri-stimulus calcium traces, and comparative violin plots. & Dynamics-Behavior-Anatomy-Gene Coupling  \\
multi-mouse-spike\_behavior\_visual\_stim-spearman-scatter\_plot & Compares baseline-corrected neurometric shifts from electrophysiology and widefield imaging across matched cortical regions using Spearman rank correlation. It produces an annotated scatter plot with a regression line, 95\% confidence interval, correlation coefficient, and p-value. & Dynamics-Behavior-Stimulus Coupling  \\
multi-macaque-spike\_behavior\_visual\_stim-latent\_dynamical\_model-log\_likelihood\_ratio & Benchmarks latent dynamical models against trial-averaged PSTH baselines using held-out marginal log-likelihood differences across neurons. It outputs the improvement distribution, mean population gain, and a Wilcoxon p-value testing whether gains exceed zero. & Dynamics-Behavior-Stimulus Coupling  \\
multi-zebrafish-gene\_anatomy-expression\_level\_clustering-heatmap & Generates a publication-quality heatmap of larval zebrafish gene expression after excluding selected brain regions and z-scoring expression across genes within each region. Outputs the heatmap and summary statistics to reveal region-specific spatial expression patterns relevant to functional connectivity. & Molecule-Structure Coupling  \\
multi-zebrafish-gene\_anatomy-expression\_count\_mapping-spatial\_density & Maps eomesa- and pvalb7-expressing zebrafish neurons by deformably registering EASI-FISH volumes to MapZebrain, assigning RNA spots to segmented cells, and thresholding expression. It produces atlas-aligned 3D point clouds and per-cell tables for individual and co-expressing populations. & Molecule-Structure Coupling  \\
multi-mouse-spike\_visual\_stim-svm-performance & Analyzes Allen Neuropixels data using responsiveness tests and cross-validated linear SVM decoding with unit-count subsampling across brain regions. It outputs regional responsive-neuron fractions, decoding accuracy statistics, null comparisons, and comparative visualizations. & Stimulus-Dynamics Coupling  \\
multi-mouse-spike\_anatomy-significant\_neuron\_fraction-atlas\_contour\_map & Analyzes Neuropixels firing-rate differences between left and right stimuli using Mann–Whitney U tests with Benjamini–Hochberg FDR correction, aggregated by Beryl brain region. It produces a right-hemisphere Swanson flatmap showing each region’s fraction of significantly modulated neurons. & Structure Function Coupling  \\
multi-zebrafish-sc\_fc\_matrix-directionality\_anova-matrix\_histogram & Classifies zebrafish region pairs as unconnected, unidirectional, or bidirectional from the directed structural connectome, pools per-larva functional connectivity by class, and compares distributions using one-way ANOVA. It outputs category matrices, FC arrays, density plots, and F and P statistics. & Structure Function Coupling  \\
multi-zebrafish-sc\_fc\_matrix-diffusion\_map-connectivity\_gradient & Computes structural and functional connectivity gradients in the larval zebrafish brain using diffusion map embedding, then correlates the first two gradients across modalities. It saves gradient and eigenvalue arrays and produces bilateral atlas overlays showing their anatomical organization. & Structure Function Coupling  \\
multi-zebrafish-sc\_fc\_matrix-graph\_predictors-fc\_variance\_explained & Computes approximately graph-theoretic predictors from zebrafish structural connectivity and ranks them by edgewise squared Pearson correlation with functional connectivity. It reports variance explained relative to undirected structural connectivity, identifying matching-index measures as the strongest predictors. & Structure Function Coupling  \\
multi-zebrafish-sc\_fc\_matrix-hungarian\_matching-gradient\_correspondence & Matches the zebrafish structural and functional connectivity gradients using all-pairs Pearson correlations and sign-invariant Hungarian assignment. It outputs a one-to-one mode mapping, reordered correlation matrix, and absolute-similarity heatmap for assessing SC–FC gradient correspondence. & Structure Function Coupling  \\
multi-zebrafish-sc\_fc\_matrix-degree\_correlation\_polyfit-scatter\_plot & Quantifies zebrafish structure–function hub coupling by correlating weighted structural degree with positive-only functional degree and fitting an ordinary least-squares line. It outputs degree arrays, correlation and regression statistics, and a scatter plot with the best-fit line. & Structure Function Coupling  \\
multi-zebrafish-structural\_functional\_matrix-graph\_predictor\_pca\_correlation-pc1\_fc\_matrix & Extracts and reshapes the leading PCA score from zebrafish structural-connectivity predictors, sign-aligns it to functional connectivity, and measures edge-wise Pearson correlation across unique region pairs. It outputs a structural PC1 matrix, coupling statistics, and a composite structure–function heatmap. & Structure Function Coupling  \\
multi-zebrafish-sc\_fc\_matrix-node\_degree\_correlation-degree\_atlas\_map & Computes weighted structural degree and absolute functional degree across filtered zebrafish brain regions, z-scores them, and maps them onto opposite hemispheres of the dorsal Mapzebrain atlas. It outputs degree arrays, an anatomical hub map, and the structural–functional degree correlation. & Structure Function Coupling  \\
multi-zebrafish-sc\_fc\_matrix-correlation\_polyfit-scatter\_plot & Creates an edge-wise scatter plot of structural PC1 versus functional connectivity and fits an ordinary least-squares linear regression, reporting Pearson r, R², slope, and intercept. It visualizes structure–function coupling and the proportion of FC variance explained by structural PC1. & Structure Function Coupling  \\
\bottomrule
\end{longtable}
\endgroup

\begingroup
\scriptsize
\setlength{\tabcolsep}{2.5pt}
\renewcommand{\arraystretch}{1.12}
\setlength{\emergencystretch}{2em}

\setlength{\LTleft}{0pt}
\setlength{\LTright}{\fill}

\begin{longtable}{
  @{}
  >{\RaggedRight\arraybackslash}p{0.235\textwidth}
  >{\RaggedRight\arraybackslash}p{0.080\textwidth}
  >{\RaggedRight\arraybackslash}p{0.042\textwidth}
  >{\RaggedRight\arraybackslash}p{0.075\textwidth}
  >{\RaggedRight\arraybackslash}p{0.048\textwidth}
  >{\RaggedRight\arraybackslash}p{0.230\textwidth}
  >{\RaggedRight\arraybackslash}p{0.260\textwidth}
  @{}
}
\caption{Source publications used in BrainArena. Only publications contributing tasks to the current 32-task benchmark are included. Num. tasks denotes the number of figure-reproduction tasks drawn from each publication.}
\label{tab:brainarena_source_publications}\\
\toprule
\textbf{Title}
& \textbf{Venue}
& \textbf{Year}
& \textbf{Organism}
& \textbf{Num. tasks}
& \textbf{Data modality}
& \textbf{Analyses} \\
\midrule
\endfirsthead

\caption[]{Source publications used in BrainArena.}\\
\toprule
\textbf{Title}
& \textbf{Venue}
& \textbf{Year}
& \textbf{Organism}
& \textbf{Num. tasks}
& \textbf{Data modality}
& \textbf{Analyses} \\
\midrule
\endhead

\midrule
\multicolumn{7}{r}{\emph{Continued on next page}}\\
\endfoot

\bottomrule
\endlastfoot

A brain-wide map of neural activity during complex behaviour & Nature & 2025 & mouse & 2 & Neuropixels electrophysiology, visual decision behavior and brain-atlas annotations & Brain-wide modulation mapping; population trajectory analysis \\
\addlinespace
Brain-wide representations of prior information in mouse decision-making & Nature & 2025 & mouse & 4 & Neuropixels electrophysiology, widefield calcium imaging and block-structured visual decision behavior & Neural prior decoding; neurometric curve construction; Bayesian model comparison; neural--behavioral decay correlation \\
\addlinespace
Plastic landmark anchoring in zebrafish compass neurons & Nature & 2026 & zebrafish & 3 & Calcium imaging of head-direction neurons in virtual-reality visual environments & Compass alignment quantification; circular tuning-curve analysis; ring-attractor phase mapping \\
\addlinespace
Brain-wide representations of behavior spanning multiple timescales and states in \textit{C.~elegans} & Cell & 2023 & \textit{C.~elegans} & 5 & Brain-wide calcium imaging and freely moving behavioral recordings & Multi-timescale filtering; timescale distribution characterization; temporal-offset decoding; manifold and PCA population analyses \\
\addlinespace
Whole-brain co-mapping of gene expression and neuronal activity at cellular resolution in behaving zebrafish & bioRxiv & 2026 & zebrafish & 3 & Whole-brain calcium imaging, multiplexed gene-expression mapping, behavior and anatomical coordinates & Molecular subtype classification; marker-gene spatial distribution; visual-response profiling \\
\addlinespace
Multimodal cue integration and learning in a neural representation of head direction & Nature Neuroscience & 2025 & \textit{Drosophila} & 2 & Population calcium imaging in multimodal virtual-reality environments & Cue-dependent encoding comparison; bump-width and encoding-accuracy correlation \\
\addlinespace
Continuous cell-type diversification in mouse visual cortex development & Nature & 2025 & mouse & 3 & Single-cell transcriptomic and epigenomic atlas of developing visual cortex & Developmental expression trajectories; cell-type diversity quantification; differential expression analysis \\
\addlinespace
The dynamics and geometry of choice in the premotor cortex & Nature & 2025 & macaque & 4 & Premotor-cortex electrophysiology during perceptual decision-making & Single-neuron PSTH characterization; spike-raster visualization; latent-dynamics model fitting; simultaneous population recording \\
\addlinespace
Cholinergic feedback for modality- and context-specific modulation of sensory representations & Science & 2025 & mouse & 2 & Cholinergic feedback recordings, olfactory-task behavior and chemogenetic manipulation & Odor-response classification; chemogenetic learning-manipulation analysis \\
\addlinespace
Structural and genetic determinants of zebrafish functional brain networks & Science Advances & 2025 & zebrafish & 4 & Whole-brain calcium imaging, structural-connectivity matrices and spatial/genetic annotations & Functional connectivity estimation; structural network visualization; community detection; spatial response mapping \\

\end{longtable}
\endgroup

 \begingroup
  \scriptsize
  \setlength{\tabcolsep}{2.5pt}
  \renewcommand{\arraystretch}{1.12}
  \setlength{\emergencystretch}{2em}
  \begin{longtable}{@{}>{\RaggedRight\arraybackslash}p{0.205\textwidth}>{\RaggedRight\arraybackslash}p{0.215\textwidth}
  >{\RaggedRight\arraybackslash}p{0.075\textwidth}>{\RaggedRight\arraybackslash}p{0.470\textwidth}@{}}
  \caption{BrainArena task metadata.}\\
  \label{tab:neurosci-bench-tasks}\\
  \toprule
  \textbf{Task Topic} & \textbf{Category} & \textbf{Difficulty} & \textbf{Algorithms / methods} \\
  \midrule
  \endfirsthead
  \caption[]{BrainArena task metadata. Continued.}\\
  \toprule
  \textbf{Task Topic} & \textbf{Category} & \textbf{Difficulty} & \textbf{Algorithms / methods} \\
  \midrule
  \endhead
  \midrule
  \multicolumn{4}{r}{Continued on next page}\\
  \endfoot
  \bottomrule
  \endlastfoot
  Multi-timescale filtering & Neural Response Characterization & Easy & EWMA filtering, Pearson correlation, Z-score normalization \\
  Timescale distribution characterization & Neural Response Characterization & Easy & ECDF computation, Median pooling, Cross-animal
  statistics (SD) \\
  Temporal offset decoding & Neural Decoding & Hard & Ridge regression (CV), Contiguous-block CV, Circular shift shuffling \\
  Neural manifold embedding & Population Representation Analysis & Easy & UMAP, Hierarchical clustering (single-linkage), Behavioral
  encoding classification \\
  Population dynamics decomposition & Population Representation Analysis & Easy & PCA, Hierarchical clustering (single-linkage),
  Weighted PC loadings \\
  \addlinespace
  Neural prior decoding & Neural Decoding & Hard & Ridge regression (CV), Bayes-optimal prior (HMM), Circular shift null \\
  Neurometric curve construction & Neural Decoding & Hard & Logistic decoder (CV), Psychometric curve fitting, Prior-conditioned
  neurometric \\
  Bayesian model comparison & Computational Modeling & Hard & Bayesian model selection (BMS/BIC), Logistic regression, Exceedance
  probability \\
  Cross-modal decay correlation & Statistical Analysis & Hard & Action kernel fitting, Exponential decay fitting, Pearson correlation \\
  \addlinespace
  Odor response classification & Statistical Analysis & Easy & Response classification (threshold-based), Fraction computation \\
  Chemogenetic learning manipulation & Statistical Analysis & Easy & Two-way ANOVA, SEM computation, Learning curve analysis \\
  \addlinespace
  Compass alignment quantification & Statistical Analysis & Hard & Population vector decoding, Circular offset calculation, Wilcoxon
  signed-rank test \\
  Circular tuning curve analysis & Neural Response Characterization & Hard & Circular histogram binning, Rayleigh test, Tuning curve
  construction \\
  Ring attractor phase mapping & Neural Response Characterization & Hard & Population vector decoding, Circular correlation, Circular
  harmonic fitting \\
  \addlinespace
  Cue-dependent encoding comparison & Statistical Analysis & Hard & Circular variance, Von Mises bump fitting, Linear mixed-effects
  model, Tukey post-hoc \\
  Bump-accuracy correlation & Statistical Analysis & Hard & Pearson correlation, Von Mises bump fitting, Linear regression \\
  \addlinespace
  Functional connectivity estimation & Brain Network \& Anatomical Mapping & Easy & Pearson correlation (pairwise), Group averaging
  (nan-mean), Inter-subject reproducibility \\
  Structural network visualization & Brain Network \& Anatomical Mapping & Hard & Matrix symmetrization, Coordinate transformation,
  Weighted network construction \\
  Community detection & Brain Network \& Anatomical Mapping & Hard & Louvain consensus clustering (iterated), Co-assignment matrix, Ward
  hierarchical clustering \\
  Spatial response mapping & Brain Network \& Anatomical Mapping & Hard & Voxel grid construction, Monte Carlo spatial threshold,
  Gaussian smoothing \\
  \addlinespace
  Molecular subtype classification & Molecular \& Developmental Analysis & Hard & Combinatorial gene expression classification, Log1p
  transformation, Cross-animal filtering \\
  Marker gene spatial distribution & Brain Network \& Anatomical Mapping & Easy & Gene expression thresholding, 3D spatial projection,
  Marker co-expression classification \\
  Visual response profiling & Population Representation Analysis & Hard & Percentile thresholding, Hierarchical clustering (response
  similarity), Multi-stimulus heatmap sorting \\
  \addlinespace
  Single-neuron PSTH characterization & Neural Response Characterization & Easy & PSTH computation, Trial averaging, SEM calculation \\
  Latent dynamics model fitting & Computational Modeling & Hard & Latent dynamical model, Poisson log-likelihood, Cross-validated model
  comparison, Wilcoxon signed-rank test \\
  Spike raster visualization & Neural Response Characterization & Easy & Spike raster construction, PSTH computation, Condition-grouped
  trial averaging \\
  Simultaneous population recording & Neural Response Characterization & Easy & PSTH computation, Multi-neuron simultaneous recording,
  Trial averaging \\
  \addlinespace
  Brain-wide modulation mapping & Brain Network \& Anatomical Mapping & Hard & Mann-Whitney U test, FDR correction, Permutation test
  (block-stratified), Swanson flatmap rendering \\
  Population trajectory analysis & Population Representation Analysis & Hard & PCA, Euclidean distance (population trajectory),
  Permutation test, Latency/amplitude extraction \\
  \addlinespace
  Developmental trajectory tracking & Molecular \& Developmental Analysis & Hard & Mean expression per age, LOWESS smoothing,
  Developmental trajectory classification \\
  Cell-type diversity quantification & Molecular \& Developmental Analysis & Easy & Unique cluster counting, Developmental time series,
  Event annotation \\
  Differential expression analysis & Molecular \& Developmental Analysis & Hard & Pseudobulk aggregation, Differential expression
  testing, FDR correction, Log2 fold-change \\
  \end{longtable}
  \endgroup

\begingroup
\footnotesize
\renewcommand{\arraystretch}{0.92}
\setlength{\tabcolsep}{4.5pt}

\setlength{\LTleft}{\fill}
\setlength{\LTright}{\fill}

\setlength{\LTcapwidth}{1\linewidth}

\begin{longtable}{
    >{\raggedright\arraybackslash}
    p{\dimexpr0.36\linewidth-2\tabcolsep\relax}
    >{\centering\arraybackslash}
    p{\dimexpr0.13\linewidth-2\tabcolsep\relax}
    >{\centering\arraybackslash}
    p{\dimexpr0.13\linewidth-2\tabcolsep\relax}
    >{\centering\arraybackslash}
    p{\dimexpr0.13\linewidth-2\tabcolsep\relax}
}

\caption{
\textbf{System-level ablation of the Co-evolving Neuroscience Analysis Repertoire on BrainArena.}
We report task-level scores of SeekBrain with and without the Repertoire.
Diff. denotes the score difference between the full system and the ablated system.
Tasks are grouped by difficulty level.
}
\label{tab:repertoire_ablation_details}\\

\toprule
\textbf{Task}
& \textbf{w/ Rep.}
& \textbf{w/o Rep.}
& \textbf{Diff.} \\
\midrule
\endfirsthead

\toprule
\textbf{Task}
& \textbf{w/ Rep.}
& \textbf{w/o Rep.}
& \textbf{Diff.} \\
\midrule
\endhead

\midrule
\multicolumn{4}{r}{\footnotesize Continued on next page} \\
\endfoot

\bottomrule
\endlastfoot

\multicolumn{4}{l}{\textbf{Easy}} \\
\midrule

    Multi-timescale filtering              & 76 & 67 & 9  \\
    Odor response classification           & 74 & 32 & 42 \\
    Chemogenetic learning manipulation     & 92 & 75 & 17 \\
    Ring attractor phase mapping            & 76 & 48 & 28 \\
    Cue-dependent encoding comparison       & 71 & 69 & 2  \\
    Bump-accuracy correlation               & 77 & 62 & 15 \\
    Functional connectivity estimation     & 92 & 89 & 3  \\
    Structural network visualization       & 89 & 56 & 33 \\
    Marker gene spatial distribution       & 81 & 67 & 14 \\
    Single-neuron PSTH characterization    & 84 & 78 & 6  \\
    Spike raster visualization             & 91 & 83 & 8  \\
    Simultaneous population recording      & 94 & 71 & 23 \\
    Cell-type diversity quantification     & 74 & 63 & 11 \\
    
\midrule
\multicolumn{4}{l}{\textbf{Hard}} \\
\midrule

    Timescale distribution characterization & 87 & 76 & 11  \\
    Temporal offset decoding                & 89 & 82 & 7   \\
    Neural manifold embedding               & 89 & 83 & 6   \\
    Population dynamics decomposition       & 72 & 53 & 19  \\
    Neural prior decoding                   & 92 & 79 & 13  \\
    Neurometric curve construction          & 84 & 83 & 1   \\
    Bayesian model comparison               & 85 & 81 & 4  \\
    Cross-modal decay correlation           & 59 & 55 & 4 \\
    Compass alignment quantification        & 74 & 51 & 23  \\
    Circular tuning curve analysis          & 66 & 51 & 15  \\
    Community detection                     & 81 & 76 & 5   \\
    Spatial response mapping                & 63 & 39 & 24  \\
    Molecular subtype classification        & 58 & 46 & 12  \\
    Visual response profiling               & 65 & 58 & 7   \\
    Latent dynamics model fitting           & 46 & 42 & 4   \\
    Brain-wide modulation mapping           & 75 & 52 & 23  \\
    Population trajectory analysis          & 58 & 57 & 1   \\
    Developmental trajectory tracking       & 46 & 42 & 4  \\
    Differential expression analysis        & 66 & 58 & 8   \\

\end{longtable}
\endgroup

\begingroup
\scriptsize
\setlength{\tabcolsep}{1.4pt}
\renewcommand{\arraystretch}{1.12}
\setlength{\emergencystretch}{2em}
\begin{longtable}{@{} p{0.25\linewidth} p{0.69\linewidth} @{}}
\caption{Representative rubric-level errors after removing the Repertoire.}
\label{tab:repertoire_rubric_errors}\\
\toprule
\textbf{Task} & \textbf{Error introduced or worsened after removing the Repertoire} \\
\midrule
\endfirsthead

\toprule
\textbf{Task} & \textbf{Error introduced or worsened after removing the Repertoire} \\
\midrule
\endhead

\midrule
\endfoot

\bottomrule
\endlastfoot

Multi-timescale filtering &
The overall analysis remained executable and retained the EWMA interpretation, but neuron selection became less faithful: all three example correlations were negative and were interpreted as locomotion suppression, whereas the source examples were positively correlated. The short, intermediate, and long half-decay estimates also moved farther from the reference values, slightly weakening both the timescale and biological-conclusion scores. \\
\addlinespace

Odor response classification &
The ablated analysis replaced the required paired pie charts with bars and produced substantially incorrect response fractions, especially an inflated suppressed fraction. It then emphasized a significant state-dependent redistribution rather than the intended similarity in the overall responsive fraction between passive and task conditions. These errors were much more severe than the pre-ablation overinterpretation of a small task-related increase. \\
\addlinespace

Chemogenetic learning manipulation &
The ablated figure collapsed the post-CNO/no-CNO recovery condition into a continuation of the inhibited-group trace instead of displaying the required third group separately, and it omitted the principal significance annotations. It also replaced the two-way group-by-day ANOVA with a Mann--Whitney test, obscuring the experimental design even though the impairment and recovery conclusions remained broadly correct. \\
\addlinespace

Ring attractor phase mapping &
The learning panel lost the slope-2 reference and was interpreted as a two-branch, 180-degree ``flip-flop'' state. This is the prediction that the experiment was meant to go beyond: the key result is a stretched mapping in which 180 degrees of scene orientation spans the full 360-degree bump phase. The ablated analysis also misplaced the twofold signature into the post-training epoch and did not clearly recover the post-learning return to the diagonal mapping. \\
\addlinespace

Cue-dependent encoding comparison &
The score changed only slightly because the pre-ablation analysis had already missed the significant dim-versus-bright comparison. After ablation, an additional methodological error appeared: HD accuracy was no longer defined as one minus circular variance (equivalently, resultant-vector length), and the pairwise comparison brackets were omitted. The erroneous conclusion that dim and bright cues do not differ therefore persisted. \\
\addlinespace

Bump-accuracy correlation &
The ablated analysis replaced HD accuracy (resultant-vector length of the circular offset) with the absolute circular--circular correlation, and its bump-width calculation yielded implausibly narrow values. Consequently the negative across-fly relationship became only marginal and non-significant ($p=0.068$), whereas the pre-ablation result recovered a significant negative correlation consistent with the intended narrower-bump/higher-accuracy conclusion. \\
\addlinespace

Functional connectivity estimation &
The matrix construction remained largely correct and the total score decreased only slightly. The principal new loss was interpretive: the ablated conclusions did not clearly explain the matrix as mesoscopic functional co-activity among larval zebrafish regions, and the figure omitted the major brain-subdivision labels. The computation of region-averaged Pearson correlations itself was preserved. \\
\addlinespace

Structural network visualization &
The ablated pipeline plotted every positive edge rather than suppressing the bottom quartile, producing a much denser and less interpretable network, and the side-view coordinate handling was inconsistent. It also failed to state clearly that edge weights estimate relative structural wiring strength. These were substantial new degradations beyond the pre-ablation presentation-level limitations. \\
\addlinespace

Marker gene spatial distribution &
The ablated maps weakened the expected rostral-telencephalic enrichment of \textit{eomesa}, used unverified hard-coded gene-column assignments, and implemented the lateral view with a questionable axis choice. The conclusions continued to omit lateral dorsal-pallium \textit{pvalb7}, Purkinje-cell identity, and hippocampal homology; thus the pre-existing anatomical omissions remained while the spatial mapping and method became less reliable. \\
\addlinespace

Single-neuron PSTH characterization &
All four ablated examples showed broadly similar rise-and-fall profiles, further reducing the intended contrast between canonical ramps and non-monotonic responses. The analysis also used 20-ms bins plus Gaussian smoothing instead of 75-ms windows advanced by 10 ms and gave a slightly less bounded heterogeneity conclusion. \\
\addlinespace

Spike raster visualization &
The ablated analysis changed the PSTH construction from the specified 75-ms sliding window to 20-ms bins with Gaussian smoothing. More importantly, its interpretation largely ignored the trial-by-trial spike-timing variability visible in the rasters, discussing reaction-time ordering without extracting the raster's main single-trial information. \\
\addlinespace

Simultaneous population recording &
Although the ablated figure retained 14 simultaneously recorded units, incompatible temporal binning/smoothing and a broader time window reduced fidelity. The largest loss was interpretive: the diverse temporal profiles were no longer clearly tied to intrinsic heterogeneity within the simultaneously recorded local PMd population, producing a much lower population-heterogeneity score. \\
\addlinespace

Cell-type diversity quantification &
The counting procedure and overall developmental trend were preserved, but the ablated interpretation became substantially less precise. It failed to describe the early decline and the multiple bouts of subcluster expansion, omitted the P20--P28 critical-period context, and suggested activity-dependent diversification without adequately maintaining the association-versus-causation boundary. \\
\addlinespace

Timescale distribution characterization &
The ablated analysis pooled neurons and bootstrapped an artificial variability band instead of computing ECDFs per animal, while also reporting 42 rather than 14 animals. Its recovered distribution was truncated at about 16.5 s, losing the long-timescale tail, and it introduced speculative higher-order/state-tracking interpretations rather than clearly stating a short-timescale majority with a long tail. \\
\addlinespace

Temporal offset decoding &
The basic decoder analysis survived, but the ablated curve became noisier, with very wide intervals and neural performance exceeding the purported best-possible reference near zero offset. The conclusion also softened the key asymmetry to ``past more strongly than future'' instead of clearly stating that future velocity is not reliably decoded above baseline. \\
\addlinespace

Neural manifold embedding &
The ablated UMAP omitted the background of unclassified neurons and appeared substantially sparser and less continuous than the pre-ablation embedding. It also became less explicit that distance in the CePNEM-derived manifold represents behavioral-encoding similarity, although the activated/inhibited gradient and continuous-landscape conclusion were retained. \\
\addlinespace

Population dynamics decomposition &
The ablated analysis no longer established that the displayed PCs were obtained from neural activity, instead taking quantities from hierarchical model parameters; it also used population-level tuning scalars rather than clearly defined per-neuron CePNEM categories. The combined heatmap lost fidelity, and the already incorrect discrete-module interpretation persisted rather than recovering complex mixtures with limited cluster structure. \\
\addlinespace

Neural prior decoding &
The ablated decoder did not clearly distinguish ordinary from pseudosession-corrected $R^2$, reported a mismatched value, and weakened the explanation of why the inter-trial window minimizes movement contamination. The scatter orientation and visual match also degraded slightly, although decoding remained correctly localized to vlOFC/ORBvl and retained the anticipatory-prior interpretation. \\
\addlinespace

Neurometric curve construction &
There was little net change. The ablated analysis simplified the Bayes-optimal prior to block probability and pooled sessions, while retaining the pre-existing failure to recover the significant dim-to-bright/zero-contrast prior shift and the emerging-choice caveat. Thus the score was nearly unchanged rather than showing a distinct new conclusion-level failure caused by ablation. \\
\addlinespace

Bayesian model comparison &
It added an acceptable companion panel and improved the framing of action-history models, while the already present limitations---incomplete model-selection details and minor imprecision about subjective priors---remained. \\
\addlinespace

Cross-modal decay correlation &
No new rubric-level error was introduced; the ablated score increased from 49 to 60. It corrected the widefield correlation's sign, but both modalities were represented by sparse or surrogate data with severely inflated correlations, incorrect statistics, and no sufficiently specific interpretation in terms of matched neural and behavioral past-action decay. \\
\addlinespace

Compass alignment quantification &
The ablated curve lost the strong learning-specific increase and post-learning decrease in out-phase time. It then attributed the change to nonspecific time-dependent degradation rather than the two-sun manipulation and experience-dependent plasticity, directly reversing the intended biological interpretation. Simplified HD-cell/bump estimation compounded this conclusion-level error. \\
\addlinespace

Circular tuning curve analysis &
The ablated heatmaps assigned bimodality to the post-training rather than learning epoch and described learning as degraded tuning. It completely omitted the ordered rotation of tuning curves as a function of pre-learning preferred orientation, so it missed the population-level, experience-dependent remapping that distinguishes the result from simple response duplication. \\
\addlinespace

Community detection &
The ablated consensus used only 200 Louvain runs rather than the source-style thousands, yielding a more uniform and overly blocky matrix. It continued to present four modules as a definitive partition and became even less explicit about multilevel hierarchical organization, although the basic consensus-matrix and dendrogram construction remained intact. \\
\addlinespace

Spatial response mapping &
The ablated maps used point displays and inconsistent normalization that obscured the source density patterns, including the left-habenula light-responsive concentration. The conclusions omitted the distributed hindbrain motor-positive pattern, the restricted motor-negative organization, accurate dark-responsive anatomy, and light-response lateralization; the underlying exponential-kernel regressor definition was also not explained. \\
\addlinespace

Molecular subtype classification &
The ablated method relied directly on a precomputed binary-gene matrix instead of constructing it from continuous counts with gene- and fish-specific thresholds, and omitted spatial filtering, cross-fish reproducibility, and cell-support criteria. It also failed to report the distribution of subtypes by number of positive genes and overemphasized broad anatomical distribution rather than non-random regional enrichment. \\
\addlinespace

Visual response profiling &
The ablated figure retained a large sorted population and improved some visual structure, but supplied essentially no scientific conclusion. It therefore failed to interpret the stimulus-locked ensembles, identify forward visual motion as the dominant swimming-coupled response, or connect molecular subtypes to distributed functional responses; first-trial exclusion and reproducible neuron ordering were also under-specified. \\
\addlinespace

Latent dynamics model fitting &
The already non-source latent model was further replaced by an incompatible latent AR(1) GLM, eliminating the remaining model-fitting/model-selection credit. Cross-validation omitted reaction-time likelihood removal and used a per-spike rather than source-scale likelihood ratio. The direction of model advantage remained correct, but medians and sample sizes were still substantially wrong. \\
\addlinespace

Brain-wide modulation mapping &
The ablated analysis shifted the response window from 0--100 ms to 50--150 ms, pooled neurons across sessions, and replaced session-level Fisher combination with a regional binomial/BH procedure at the wrong threshold. It consequently reported 34/223 rather than 19/201 significant regions and omitted both the concentration in named visual cortical/thalamic areas and significant non-classical regions. \\
\addlinespace

Population trajectory analysis &
The overall score was nearly unchanged, but the ablated trajectories had non-zero baselines and non-standard normalization. Its latency permutation did not fully preserve correlated task strata, it placed LP too late, and it omitted VISpm/VISam and the later significant midbrain/hindbrain sequence; thus the intended region-by-region temporal hierarchy remained incomplete. \\
\addlinespace

Developmental trajectory tracking &
Both versions omitted prenatal-antecedent mapping and GAM smoothing and under-described within-class similarity, Tbr1 layer divergence, L5 ET downregulation, and complex transcription-factor trajectories; the ablated version modestly improved figure structure without resolving these persistent errors. \\
\addlinespace

Differential expression analysis &
The ablated analysis retained the pre-existing wrong developmental windows and non-source Wilcoxon workflow, but its biological interpretation worsened: it characterized deep-layer glutamatergic neurons as downregulation-dominant, contrary to the expected neuronal upregulation bias. It also failed to generalize downregulation across non-neuronal subclasses, missed the strong microglial effect, and weakened the causation boundary. \\

\end{longtable}
\endgroup

\begingroup
\footnotesize
\renewcommand{\arraystretch}{0.92}
\setlength{\tabcolsep}{2.5pt}
\setlength{\LTleft}{\fill}
\setlength{\LTright}{\fill}
\setlength{\LTcapwidth}{1\linewidth}

\begin{longtable}{@{}>{\raggedright\arraybackslash}p{\dimexpr0.75\linewidth-2.2cm-2\tabcolsep\relax}>{\centering\arraybackslash}p{2.2cm}@{}}
\caption{
\textbf{Brain region name.} The full names of the brain region abbreviations in Figs. 3 and 4.
}
\label{tab:brain_region_abbr} \\
\toprule
\textbf{Brain region} & \textbf{Abbreviation} \\
\midrule
\endfirsthead

\caption[]{\textbf{Brain region name.} (continued)} \\
\toprule
\textbf{Brain region} & \textbf{Abbreviation} \\
\midrule
\endhead

\midrule
\multicolumn{2}{r}{\footnotesize Continued on next page} \\
\endfoot

\bottomrule
\endlastfoot

\multicolumn{2}{@{}l}{\textbf{Zebrafish brain regions}} \\
\midrule
Olfactory Bulb & OB \\
Pallium & P \\
Subpallium & S \\
Preoptic Region & Po \\
Retina & R \\
Optic Tract and Accessory Optic System & Otr \\
Epiphysis (Pineal gland) & Ep \\
Habenula & H \\
Pretectum & Pr \\
Migrated posterior tubercular region & mPT \\
Dorsal Thalamus & dTh \\
Ventral Thalamus & vTh \\
Posterior Tuberculum & PT \\
Rostral Hypothalamus & rH \\
Diffuse Nucleus of the Intermediate Hypothalamus & diH \\
Intermediate Hypothalamus & iH \\
Caudal Hypothalamus & cH \\
Pituitary & Pit \\
Torus Longitudinalis & TL \\
Neuropil Layers (optic tectum) & Np \\
Periventricular Cell Layer (optic tectum) & Ot \\
Torus Semicircularis & TS \\
Tegmentum & T \\
Nucleus of Medial Longitudinal Fascicle & nMLF \\
Oculomotor Nerve Nucleus & nIII \\
Trochlear Nerve Motor Nucleus & nIV \\
Interpeduncular Nucleus & IPN \\
Superior Raphe & sR \\
Valvula Cerebelli & VC \\
Corpus Cerebelli & CC \\
Eminentia Granularis & EG \\
Medial octavolateralis nucleus & MON \\
Rhombomere 1 & R1 \\
Rhombomere 2 & R2 \\
Rhombomere 3 & R3 \\
Inferior Raphe & iR \\
Rhombomere 4 & R4 \\
Medial Vestibular Nucleus & mVe \\
Facial nerve motor nucleus and octavolateralis efferent neurons & nVII' \\
Facial nerve motor nucleus and octavolateralis efferent neurons & nVII \\
Rhombomere 5 & R5 \\
Rhombomere 6 & R6 \\
Rhombomere 7 & R7 \\
Glossopharyngeal Vagal nerve motor nucleus & nX \\
Rhombomere 8 & R8 \\
Vertebra 01 ventral lateral grey matter & lV01 \\
Vertebra 01 white matter & wV01 \\
Olfactory Epithelium & OE \\
Trigeminal Ganglion & gT \\
Anterior Lateral Line Ganglion & gaLL \\
Facial Ganglion & gF \\
Octaval ganglion & gO \\
\midrule
\multicolumn{2}{@{}l}{\textbf{Mouse brain regions in the Allen Common Coordinate Framework template}} \\
\midrule
gigantocellular reticular nucleus & GRN \\
magnocellular reticular nucleus & MARN \\
lateral preoptic area & LPO \\
posterior amygdalar nucleus & PA \\
central lateral nucleus of the thalamus & CL \\
main olfactory bulb & MOB \\
facial motor nucleus & VII \\
substantia nigra, reticular part & SNr \\
primary somatosensory area, nose & SSp-n \\
visceral area & VISC \\
olivary pretectal nucleus & OP \\
pontine reticular nucleus, caudal part & PRNc \\
intertrigeminal nucleus & I5 \\
primary auditory area & AUDp \\
inferior colliculus & IC \\
nucleus of the brachium of the inferior colliculus & NB \\
anterolateral visual area & VISal \\
\end{longtable}
\endgroup

\end{document}